\documentclass[10pt, journal, twoside]{IEEEtran}

\normalsize

\usepackage{amsmath}
\usepackage{amssymb}
\usepackage{amsfonts}
\usepackage{graphicx}
\usepackage{epstopdf}
\usepackage{subfigure} 
\usepackage[dvipsnames]{xcolor}
\usepackage{bm}
\usepackage{url}
\usepackage[ruled,boxed,linesnumbered,commentsnumbered]{algorithm2e}
\usepackage{setspace}
\usepackage{cite}
\usepackage{upgreek}
\usepackage[colorlinks,linkcolor=blue,anchorcolor=blue,citecolor=blue]{hyperref} 

\usepackage{amsthm}

\newtheorem{remark}{Remark}

\newcommand{\algref}[1]{\textbf{Algorithm \ref{#1}}}

\newcommand{\figref}[1]{Fig.~\ref{#1}}
\newcommand{\secref}[1]{Section~\ref{#1}}

\allowdisplaybreaks[4]

\ifCLASSINFOpdf
\else
\fi

\begin{document}

\title{Toward Multi-Satellite Cooperative Transmission: A Joint Framework for CSI Acquisition, Feedback, and Phase Synchronization}

\author{\normalsize{Yiming~Zhu,~\IEEEmembership{Graduate Student Member,~IEEE}, 
Yafei~Wang,~\IEEEmembership{Graduate Student Member,~IEEE}, 
Carla~Amatetti,~\IEEEmembership{Member,~IEEE}, 
Alessandro~Vanelli-Coralli,~\IEEEmembership{Senior Member, IEEE}, 
Wenjin~Wang,~\IEEEmembership{Member,~IEEE}, 
Rui~Ding, 
Symeon~Chatzinotas,~\IEEEmembership{Fellow, IEEE}, 
Bj{\"{o}}rn~Ottersten,~\IEEEmembership{Fellow, IEEE}
} 
\thanks{Manuscript received xxx (Corresponding author: Wenjin Wang.)}
\thanks{Yiming~Zhu, Yafei~Wang, Wenjin~Wang are with the National Mobile Communications Research Laboratory, Southeast University, Nanjing 210096, China, and also with Purple Mountain Laboratories, Nanjing 211100, China (e-mail: \{ymzhu, wangyf, wangwj\}@seu.edu.cn).}
\thanks{Carla~Amatetti and Alessandro~Vanelli-Coralli are with the Department of Electrical, Electronic, and Information Engineering, University of Bologna, Bologna 40136, Italy (e-mail: \{carla.amatetti2, alessandro.vanelli\}@unibo.it).} 
\thanks{Rui Ding is with China Satellite Network Group Company Ltd., Beijing 100029, China (e-mail: greatdn@qq.com)}
\thanks{Symeon Chatzinotas and Bjorn Ottersten are with the Interdisciplinary Centre for Security, Reliability and Trust (SnT), University of Luxembourg,  (e-mail: \{symeon.chatzinotas, bjorn.ottersten\}@uni.lu).}
}

\markboth{\:}
{Submitted paper}
\maketitle

\begin{abstract}
The stringent link budget, caused by long satellite-to-ground distances and payload constraints, poses a fundamental bottleneck for single-satellite transmission performance. 
Although the advent of low Earth orbit mega-constellations makes multi-satellite cooperative transmission (MSCT), such as distributed precoding (DP), increasingly feasible, its promised cooperative gains critically rely on stringent time-frequency-phase synchronization (TFP-Sync), which is difficult to maintain under rapid channel variation and substantial feedback latency. 
To address this issue, this paper proposes a joint channel state information (CSI) acquisition, feedback, and phase-level synchronization (JCAFPS) framework for MSCT. 
Specifically, to enable reliable and overhead-efficient CSI acquisition, we first design a beam-domain adjustable phase-shift tracking reference signal transmission scheme, along with design criteria for the TRS and CSI-feedback periods. 
Then, by exploiting deterministic orbital motion and dominant line-of-sight propagation, we establish a polynomial model for the temporal evolution of delay and Doppler shift, and derive an orthogonal frequency division multiplexing-based multi-satellite signal model under non-ideal synchronization. 
The analysis reveals that, unlike the regular single-satellite case, the composite multi-satellite channel exhibits nonlinear time-frequency-varying phase behavior, which necessitates symbol- and subcarrier-wise phase precompensation for coherent transmission. 
Based on these results, we develop a practical closed-loop realization that integrates single-TRS-based channel parameter estimation, multi-TRS-based channel prediction, predictive CSI feedback, and user-specific TFP precompensation. 
Numerical results demonstrate that the proposed framework achieves accurate CSI acquisition and precise TFP-Sync, enabling DP-based dual-satellite cooperative transmission to approach the theoretical $\mathbf{6}$ dB power gain over single-satellite transmission, while remaining robust under extended prediction durations and enlarged TRS periods. 

\end{abstract}

\vspace{-1mm}
\vspace{-1mm}

\IEEEpeerreviewmaketitle

\vspace{-1mm}
\section{Introduction}
\vspace{-1mm} 
 
As one key direction for the sixth-generation technology evolution, satellite communication (SatCom) is anticipated to integrate with terrestrial networks (TNs) for ubiquitous connectivity \cite{ntontin2025vision}. 
Accordingly, mobile satellite Internet, dedicated to delivering broadband access to handheld user equipments (UEs) anytime and anywhere, has attracted considerable attention from both academia and industry \cite{wang2025Toward,38.811,38.821}. 
However, realizing such broadband access to handheld UEs is challenged by a confluence of link-budget-limiting factors: 1) large path loss induced by the long distance, 2) stringent size, weight, and power (SWaP) constraints on the payload, and 3) the compact form factor of handheld terminals that restricts the achievable antenna gain and transmit power \cite{heo2023mimo}. 
Consequently, the link budget of mobile satellite Internet remains tight and is often throughput-limiting, necessitating the development of innovative techniques enabling improved system performance~\cite{wang2025Toward}. 

To mitigate link-budget limitations, precoding serves as an effective enabler in SatCom~\cite{al2022survey,wang2024towards}. 
Although pattern-oriented precoding, including Earth-moving and Earth-fixed designs, is prevalent in deployed systems due to computational efficiency and relaxed channel state information (CSI) requirements, it is hampered by limited flexibility in interference management, which results in low spectral efficiency \cite{wang2024towards}. 
Conversely, with the advancing computational capabilities of on-board processors, user-specific precoding schemes, such as maximum ratio transmission, zero forcing, and minimum mean squared error, constitute a more competitive alternative~\cite{you2020massive,wang2025statistical,wang2025MSMS}. 
These techniques exploit user CSI, e.g., angle information, to dynamically direct high-gain beams toward spatially distributed UEs while mitigating interference, thus enhancing both signal power and system capacity. 
However, the highly dynamic nature of low Earth orbit (LEO) satellites exacerbates the channel aging, thus compromising user-specific precoding performance \cite{you2020massive}. 
Furthermore, while the proliferation of LEO mega-constellations establishes the foundation for multi-satellite cooperative transmission (MSCT) paradigms to transcend single-satellite performance ceilings \cite{abdelsadek2022distributed,xu2024enhancement,wu2025distributed,wang2026Multi}, the realization of such gains is contingent upon stringent time-frequency-phase synchronization~(TFP-Sync). 
Against this~backdrop, this paper investigates a joint CSI acquisition, feedback, and phase-level synchronization (JCAFPS) framework for MSCT, offering practical solutions for improving the user peak rate.

\subsection{Related Works and Motivations}
To improve the CSI timeliness, existing studies on single-satellite transmission have investigated channel prediction methods that exploit the temporal correlation inherent in successive channel estimates~\cite{zhang2022deep2,ying2024deep}. 
Nevertheless, a fundamental~bottleneck remains: even with perfect CSI, the precoding power gain is physically bounded by single-satellite antenna aperture. 
Since engineering feasibility and cost considerations limit the scaling of array size, the attainable gain may still be insufficient to close the link-budget deficit for mobile satellite Internet. 
Enabled by recent advances in LEO manufacturing and reusable launch vehicles, mega-constellations have rendered MSCT increasingly viable. 
Distributed precoding (DP) has emerged as a compelling solution to overcome the single-satellite bottleneck. 
With data sharing via inter-satellite links (ISLs), cooperative satellites may simultaneously transmit identical data streams to UEs by coherent precoding design. 
Crucially, coherent superposition unlocks a theoretical received power gain that scales quadratically with the number of cooperative satellites under perfect synchronization. 
Prior works in \cite{abdelsadek2022distributed,xu2024enhancement} investigated distributed multiple input multiple output (MIMO)-based satellite network design and provided spectral efficiency analyses to quantify DP gains.   

Nevertheless, the realization of such cooperative gains hinges on rigorous TFP-Sync among cooperative satellites.  
The rapid dynamics and propagation latency of multi-satellite systems significantly complicate CSI acquisition, inevitably hampering accurate synchronization \cite{yue2022collaborative}. 
In \cite{wu2025distributed}, a rigorous analysis was provided to characterize how imperfect synchronization degrades the achievable rate of DP-based MSCT.
To address this issue, the approach in \cite{ouyang2025novel} relied on the assumption of perfect position and trajectory knowledge to execute geometric-based phase precompensation and satellite-UE association to harvest cooperative gains. 
Given that perfect position and trajectory information is typically unavailable in practice, recent efforts have focused on multi-satellite robust precoding under imperfect synchronization. 
To mitigate timing impairments, the authors in \cite{chen2024asynchronous} proposed a delay estimation scheme and derived a timing-robust cooperative beamforming algorithm. 
Furthermore, considering residual delay- and Doppler-induced impairments, the authors in \cite{wang2025statistical,wang2025MSMS} investigated  
statistical CSI-based DP design and multi-stream
beamspace transmission for orthogonal frequency division multiplexing (OFDM)-based multi-satellite systems.  

The achievable gains of MSCT depend on accurate TFP-Sync, and synchronization imperfections can substantially compromise these gains.  
For coherent transmission, inter-satellite phase offsets hinder coherent combination and even yield destructive interference. 
Despite the exploration of robust precoding designs under imperfect synchronization, the pivotal challenge of TFP-Sync tailored for MSCT has yet to be adequately addressed. 
While an alternative paradigm is predicated on high-precision real-time position to execute TFP-Sync, the associated physical constraints are substantial. Considering that a carrier frequency of $2$ GHz corresponds to a wavelength of $0.15$ m, the requisite centimeter-level position accuracy entails prohibitive implementation costs \cite{zangenehnejad2021gnss}, particularly in the context of LEO mega-constellations.  
These observations highlight a fundamental and timely research problem: \textit{How to design an efficient JCAFPS framework to enable TFP-Sync for~MSCT?}

\subsection{Main Contributions}
Motivated by the considerations outlined above, this paper aims to design an efficient JCAFPS framework tailored for MSCT. Our contributions are summarized as follows: 
\begin{itemize} 
  \item We propose a comprehensive JCAFPS framework for MSCT, in which CSI acquisition, predictive feedback, and precompensation are tightly integrated into a closed loop for enabling stringent TFP-Sync. To support reliable downlink CSI acquisition under severe propagation loss and multi-satellite asynchronization, we further design a dedicated tracking reference signal (TRS) transmission mechanism tailored to satellite-to-ground (S2G) channels. Specifically, by introducing beam position (BP)-specific beam-domain mapping, TRS broadcast area (TBA)-specific precompensation, and adjustable phase-shift design for TRS, the proposed design improves CSI acquisition reliability while reducing pilot overhead. Moreover, we analyze the design criteria of the TRS period and CSI feedback period, thereby revealing the fundamental trade-off among signaling overhead, prediction validity, and phase-unwrapping reliability. 
  \item We establish a time-varying S2G channel model characterizing the dynamic evolution of S2G channel parameters. Capitalizing on the deterministic satellite motion and line-of-sight (LoS) dominance, we model the propagation delay and Doppler shift via time polynomials and elucidate the qualitative relationship between polynomial order and valid time scale. 
  Based on this characterization, we further derive an OFDM-based multi-satellite signal model with imperfect synchronization and obtain the corresponding equivalent TF-domain channel representation. Moreover, we analytically reveal that, in contrast to the regular single-satellite case, the composite multi-satellite channel exhibits nonlinear TF-varying phase behavior, necessitating the fine-grained symbol- and subcarrier-wise phase precompensation for coherent MSCT. 
  \item We develop a complete algorithmic realization of the proposed JCAFPS framework. For single-TRS processing, we devise an TRS-aided channel parameter estimation scheme that combines initial channel estimation, super-resolution delay-domain parameter extraction, and coarse Doppler shift estimation. Building on successive TRS observations, we further propose a multi-TRS-based channel prediction method based on Doppler-assisted cross-TRS phase unwrapping (PU) and polynomial coefficient estimation, through which the temporal evolution of channel parameters can be compactly represented and predicted. 
  On this basis, the UE feeds back predictive CSI to the cooperative satellites, enabling UE-specific TFP precompensation for coherent transmission. 
  Numerical results further verify that the proposed framework effectively supports the gain realization of MSCT. 
\end{itemize} 

This paper is organized as follows: 
\secref{sec:framework overview} presents the framework overview. 
\secref{sec: mSat signal model} develops the multi-satellite signal model. 
\secref{sec: JCAFPS} proposes the detailed design of the JCAFPS framework.  
Simulation results are provided in \secref{sec:simulation results}, and conclusions are drawn in \secref{sec:conclusion}.  

\textit{Notations}: 
Lowercase, bold lowercase, and bold uppercase represent scalars, column vectors, and matrices, respectively. 
The superscripts $ (\cdot)^{T} $, $ (\cdot)^{*} $, $ (\cdot)^{H} $, $ (\cdot)^{-1} $, and $ (\cdot)^{\dagger} $  denote transpose, conjugate, conjugate-transpose, inverse, and pseudoinverse. 
$ \mathbb{R} $, $ \mathbb{C} $, $ \mathbb{Z} $, and $\mathbb{B}$ denote the real, complex, integer, and binary sets. 
$[\cdot]_{i}$ and $[\cdot]_{i,j}$ denote the $i$-th vector element and the $(i,j)$-th matrix element. 
For an index set $ \mathcal{S} $ (ascending order), 
$ [\cdot]_{\mathcal{S}} $, $ [\cdot]_{:,\mathcal{S}} $, and $ [\cdot]_{\mathcal{S},:} $ represent the selected elements, columns, and rows. 
$ \Vert \cdot \Vert_{1} $, $ \Vert \cdot \Vert_{2} $, and $ \Vert \cdot \Vert_{\rm F} $ denote the $ \ell_{1} $, $ \ell_{2} $, and Frobenius norms. 
$ \mathbf{I}_{M} $ represents the $ M\times M $ dimensional identity matrix.  
$ \mathbf{1}_{N} $ and $ \mathbf{0}_{N} $ denote the $ N $ dimensional all-one and all-zero vectors. 
$ \mathsf{diag}\lbrace\cdot\rbrace $, $ \mathsf{E}\lbrace\cdot\rbrace $, and $ \mathsf{min}\lbrace\cdot\rbrace $ represent the diagonal, expectation, and minimum operators.
$ \mathsf{CN}(x;\mu,\tau) $ denotes the circularly symmetric complex Gaussian distribution of variable $ x $ with mean $ \mu $ and variance $ \tau $. 
$ \otimes $, $ \odot $, and $ \circ $ denotes the Kronecker, Hadamard, and Khatri-Rao products. 
$ \bar{\jmath}=\sqrt{-1} $ and $ \delta(\cdot) $ is the Dirac delta function.





\begin{figure}[t]
  \centering
  \includegraphics[width=1.0\linewidth]{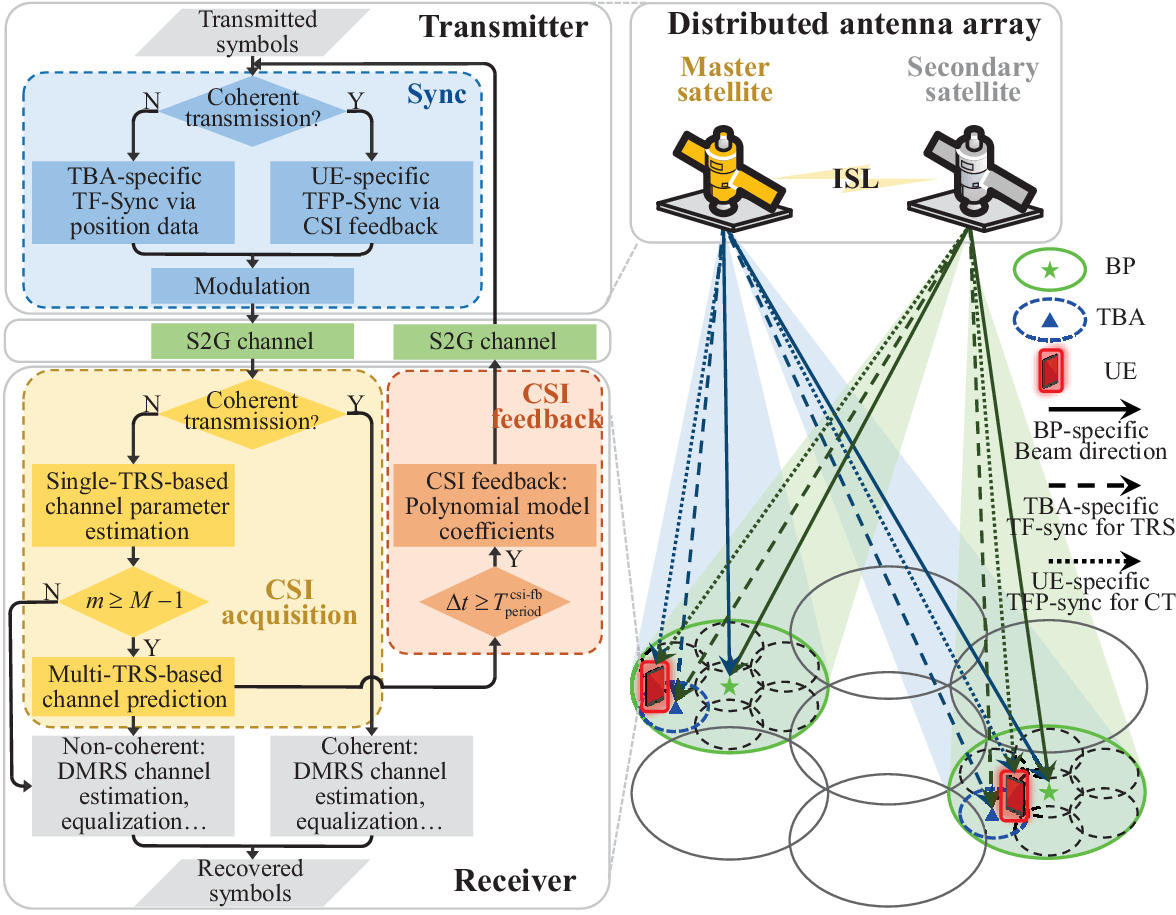}
  \vspace{-5mm}
  \caption{Flow chart of JCAFPS framework.}
  \label{fig:flow chart}
  \vspace{-2mm}
\end{figure}

\vspace{-2mm}
\section{Framework Overview}
\label{sec:framework overview}

This paper considers a frequency-division duplexing (FDD) OFDM-based multi-satellite communication system. 
As depicted in \figref{fig:flow chart}, multiple LEO satellites equipped with a uniform planar array cooperatively serve multiple single-antenna UEs over shared frequency-domain resources via space division multiple access within a common coverage area. 
Each array consists of $N_{\rm t}=N_{\rm x}N_{\rm y}$ antennas arranged in an $N_{\rm x}\times N_{\rm y}$ rectangular lattice with half-wavelength spacing. 
Owing to the non-uniform UE distribution, the service area is partitioned into $ \bar{B} $ active predivided BPs. 
To satisfy the coverage demand while effectively mitigating inter-beam interference under onboard power and hardware constraints, the satellites employ beam hopping \cite{lin2022multi}, such that $ \bar{B} $ active BPs are served over $ N_{\rm hop} $ hops and each satellite simultaneously illuminates $ B $ BPs. 
Without loss of generality, we focus on the transmission under the pattern of the first hop, as the transceiver processing remains identical across hops. 
In particular, $ S $ satellites cooperatively serve $ \bar{U} $ UEs within the $ B $ illuminated BPs, where one satellite is designated as the ``master'' to coordinate cooperative transmission \cite{38.821} and the remaining satellites act as ``secondary'' ones.  

The OFDM system comprises $ N $ subcarriers with spacing $ \Delta f $ and a cyclic prefix (CP) of length $ N_{\rm cp} $, among which $N_{\rm sc} $ subcarriers are occupied for transmission. 
Accordingly, the sampling interval, CP duration, symbol durations without and with CP, and slot duration are defined by $ T_{\rm s} = \frac{1}{N\Delta f} $, $ T_{\rm cp} = N_{\rm cp}T_{\rm s} $, $ T = NT_{\rm s} $, $ T_{\rm sym} = T_{\rm cp} + T $, and $ T_{\rm slot} $, respectively. 
The carrier frequency is $ f_{\rm c} $, corresponding to the wavelength $ \lambda = v_{\rm c}/f_{\rm c} $, where $ v_{\rm c} $ is the speed of light.

\subsection{Overall Framework Design}
\label{sec: framework}

To enable MSCT, we develop a comprehensive JCAFPS framework consisting of three coupled modules, as illustrated in \figref{fig:flow chart}. 
The framework follows a two-phase operation. 
During TRS transmission, cooperative satellites apply TBA-specific time-frequency (TF) precompensation to facilitate reliable CSI acquisition, with each TBA being a fine-grained spatial partition within a BP. 
Based on the received TRSs, the UE estimates the downlink channel parameters and progressively establishes channel prediction capability. 
During subsequent coherent transmission, the satellites exploit the fed-back predictive CSI to perform UE-specific TFP precompensation, thereby enabling the constructive superposition of identical streams at the UE\footnote{The proposed framework is applicable to MSCT with synchronization requirements, rather than being limited to coherent transmission.}. 
In this way, CSI acquisition, CSI feedback, and synchronization form a closed loop that supports the stringent TFP-Sync required by MSCT. 

\subsubsection{CSI Acquisition Module} 

Cooperative satellites periodically transmit TBA-specific TRSs, from which the UE estimates the channel parameters of different satellite-to-UE links, including delay, Doppler shift, phase, and gain. 
Owing to the deterministic orbital motion and the dominant LoS propagation of S2G channels \cite{38.811}, these parameters exhibit structured temporal evolution. The UE therefore further represents the channel dynamics using a compact set of coefficients extracted from historical estimates, thereby establishing the basis for subsequent channel prediction and synchronization. 

The main challenge is to achieve accurate multi-satellite parameter acquisition with affordable pilot overhead. This challenge stems from severe propagation loss, inter-satellite asynchronization, and the limited efficiency of conventional single-port TRS signaling in the multi-satellite setting. These issues motivate the dedicated TRS transmission and channel estimation designs presented in \secref{sec:TRS design} and \secref{sec:channel estimation}. 

\subsubsection{CSI Feedback Module} 
After sufficient TRS observations have been accumulated, the UE periodically feeds back the estimated channel-evolution coefficients to the master satellite, which then relays them to the secondary satellites through ISLs\footnote{This work focuses on downlink transmission, assuming that the reliability of uplink feedback channel is guaranteed by robust transmission strategies, such as conservative modulation and coding schemes. Detailed uplink optimization remains beyond the current scope and is reserved for future research.}. 
Based on this predictive feedback, the cooperative satellites can infer the future channel evolution and prepare the precompensation required for coherent transmission. Accordingly, the feedback module bridges UE-side channel prediction and transmitter-side precompensation. 

The key issue in this module is CSI aging caused by the substantial round-trip latency. Accordingly, the feedback should characterize channel evolution rather than merely instantaneous CSI, while keeping the signaling overhead well controlled. This motivates the model-based predictive feedback design developed in \secref{sec:trs design} and \secref{sec:channel prediction}. 

\subsubsection{Synchronization Module} 

The synchronization module supports both TRS transmission and coherent transmission. 
In the TRS phase, cooperative satellites perform open-loop TBA-specific TF precompensation based on ephemeris and TBA-center position information. 
Since the UE shares the same trajectory and position information \cite{xiang2024massive,chen2021system} and knows the adopted precompensation rule, it can reconstruct the transmitter-side compensation and recover the continuous temporal trajectories of the channel parameters. 
In the coherent transmission phase, the satellites further perform closed-loop UE-specific TFP precompensation based on the predictive feedback so as to ensure coherent combining at the UE. 

The fundamental challenge is that synchronization errors are jointly coupled across the time, frequency, and phase domains, and even small mismatches can substantially degrade coherent combining. This observation motivates the subsequent phase-aware signal modeling and predictive synchronization design in \secref{sec:phase analysis} and \secref{sec:channel prediction}.  

\vspace{-1.5mm}
\subsection{Tracking Reference Signal Transmission Design}
\label{sec:TRS design} 

\begin{figure}[t]
  \centering
  \includegraphics[trim=0mm 0.5mm 0mm 0.5mm, clip, width=1.0\linewidth]{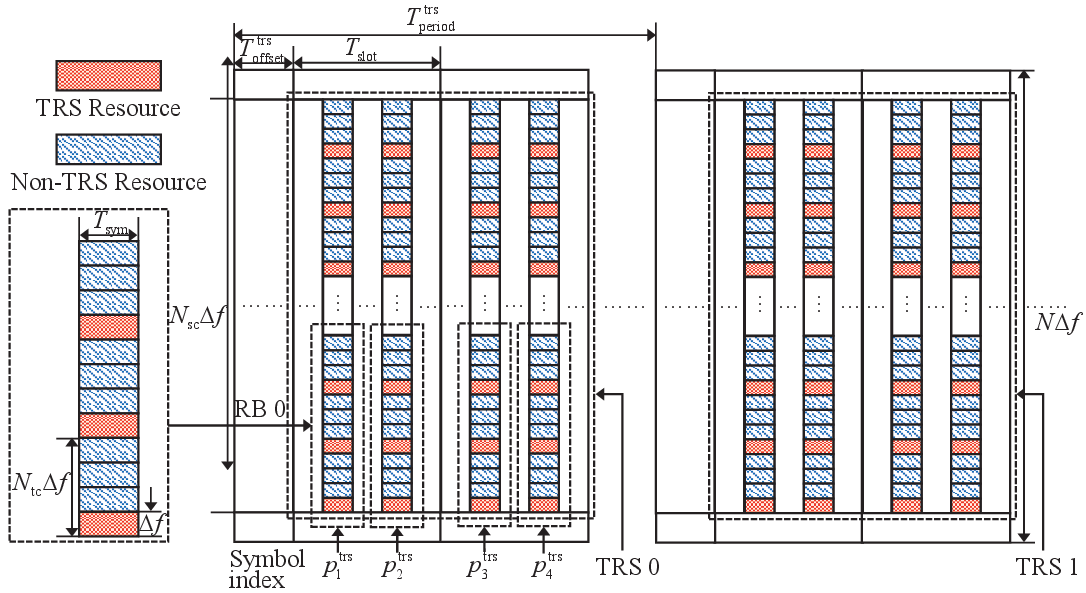}
  \vspace{-6.5mm}
  \caption{TRS configuration with $ N_{\rm sym}^{\rm trs} = 4 $, $ N_{\rm slot}^{\rm trs} = 2 $, and $ N_{\rm tc} = 4 $ \cite{38.214}.}
  \label{fig:trs configure}
  \vspace{-2mm}
\end{figure} 

MSCT requires accurate downlink channel parameter estimation to enable stringent TFP-Sync among cooperative satellites. In FDD systems, downlink CSI is typically acquired at the UE based on CSI-reference signals (CSI-RSs) and fed back via the uplink~\cite{38.211}. However, the rapid time variation of S2G channels calls for continuous high-fidelity CSI acquisition and feedback, which is not adequately supported by conventional designs. We therefore employ the TRS, defined as a set of CSI-RS resources, to support the JCAFPS framework. 
As shown in \figref{fig:trs configure}, one typical TRS configuration in frequency range~$1$ consists of $ N_{\rm sym}^{\rm trs} $ single-port periodic CSI-RS resources spanning $ N_{\rm slot}^{\rm trs} $ consecutive slots \cite{38.214}. 
The configuration is specified by the period $ T_{\rm period}^{\rm trs} $ and offset $ T_{\rm offset}^{\rm trs} $, with symbol-index set within one period denoted by $ \mathcal{N}_{\rm sym}^{\rm trs} = \{ p_{i}^{\rm trs} \}_{i=0}^{N_{\rm sym}^{\rm trs}-1} $. 
In the frequency domain, each TRS symbol occupies $ N_{\rm trs} = \frac{N_{\rm sc}}{N_{\rm tc}} $ subcarriers, defined by the index set $ \mathcal{N}_{\rm sc}^{\rm trs}$ and comb number $ N_{\rm tc} $. 
The temporal and spectral density of the TRS determine the maximum trackable frequency and timing offsets. 

To enable efficient CSI acquisition for MSCT, we design the TRS transmission tailored to the S2G channel characteristics. 

\subsubsection{BP-Specific Beam-Domain TRS Broadcasting}
To combat severe propagation loss inherent in SatCom, the TRS is mapped to the beam domain rather than the antenna domain. 
To avoid the excessive pilot overhead, each satellite steers its $ B $ simultaneous beams to the centers of predefined BPs and broadcasts TRSs to the UEs within the illuminated BPs, as shown in \figref{fig:flow chart}. 
Note that inter-beam interference can be effectively mitigated through techniques such as beam hopping, eliminating the reliance on UE-specific CSI. 

\subsubsection{TBA-Specific TRS Precompensation}
The spatial resolution of downlink beams is physically constrained by the satellite antenna aperture. With limited array sizes, the resulting BP radius may be sufficiently large to induce noticeable geometric disparities in propagation delay and Doppler shift between the BP center and edge. 
Direct precompensation targeting the BP center would thus expose edge users to multi-satellite asynchronous interference. 
To circumvent this, we propose a fine-grained approach by subdividing each BP into multiple TBAs and applying distinct TF precompensation to the transmission of each TBA, as depicted in \figref{fig:flow chart}. 
Since the number of active TBAs within a BP is typically only a small subset of the entire TBA set, the TRS transmissions associated with different active TBAs within a hop are orthogonalized in the time domain. 
TF precompensation is then tailored to the center of each active TBA, ensuring that timing and frequency offsets are strictly confined within the CP and the allowable tolerance\footnote{With the normalized frequency offset $ \epsilon $ defined relative to $\Delta f$, numerical analysis shows that $ |\epsilon| = 0.05 $ yields a signal-to-interference ratio of $ \approx 20 $ dB\cite{moose2002technique,armstrong2002analysis}. In power-limited SatCom scenarios, this implies that the impact of inter-carrier interference on link performance is negligible for $ |\epsilon| \le 0.05 $.}\cite{38.101-1,moose2002technique,armstrong2002analysis}. 
Given that the signal processing logic is generic across TBAs, the subsequent analysis focuses on the first TBA within one hop without loss of generality. 
Specifically, we consider $S$ satellites serving $U$ UEs distributed across $B$ TBAs, where each TBA is associated with a distinct BP, and all links share the same TF resources.  

\begin{remark}
As satellite payloads evolve toward extremely large-scale antenna arrays, the BP radius will progressively shrink. 
In this regime, a BP may eventually coincide with a single TBA, thereby eliminating the need for time-division multiplexing of TRS resources across TBAs.  
\end{remark}

\subsubsection{APS-TRSs for Multi-Satellite CSI Acquisition}
\label{sec:APSP} 

Since a TRS is confined to a single antenna port~\cite{38.214}, multi-satellite CSI acquisition requires orthogonal TF resources across satellites and thus incurs high pilot overhead. We therefore propose adjustable phase-shift TRSs (APS-TRSs) by leveraging the sparse delay profile of S2G channels~\cite{fontan2001statistical,vojcic2002performance,38.811}. 
Specifically, all cooperative satellites transmit TRSs on identical TF resources, while being distinguished by unique frequency-domain phase shifts. 
As a frequency-domain linear phase shift is equivalent to a delay-domain offset, the UE can resolve individual satellite channels without additional resource overhead. 
The proposed APS-TRS from the $ s $-th satellite to the $ b $-th TBA on the $ p $-th symbol of the $ m $-th TRS transmission is given~by
\begin{align}
  \label{eq:APS-TRS}
  \mathbf{d}_{sbp}^{{\rm trs},(m)} = \mathsf{diag}\{ \mathbf{f}_{\phi_{s}}^{\rm trs} \}\bar{\mathbf{d}}_{bp}^{{\rm trs},(m)},
\end{align} 
where $ \bar{\mathbf{d}}_{bp}^{{\rm trs},(m)}\in\mathbb{C}^{N_{\rm trs}} $, satisfying $ \bar{\mathbf{d}}_{bp}^{{\rm trs},(m)} (\bar{\mathbf{d}}_{bp}^{{\rm trs},(m)})^{H} = \mathbf{I}_{N_{\rm trs}} $, denotes the basic TRS sequence shared by all cooperative satellites for the $ b $-th TBA. 
$ \mathbf{f}_{\phi_{s}}^{\rm trs} $ represents the satellite-specific phase-shift vector with $ [\mathbf{f}_{\phi_{s}}^{\rm trs}]_{k} = e^{-\bar{\jmath}2\pi\frac{k\phi_{s}}{N_{\rm trs}}} $ and $ \textstyle \phi_{s} = (s+0.5)N_{\rm trs}/S $. 
This design confines the equivalent delay-domain channel after applying $ \phi_{s} $ to the interval $ \mathcal{T}_{s} =  [s/(SN_{\rm tc}\Delta f),(s+1)/(SN_{\rm tc}\Delta f)) $. 

\vspace{-3mm}
\subsection{TRS and CSI Feedback Period Design}
\label{sec:trs design}

To enable efficient TRS-aided downlink CSI acquisition and feedback without incurring excessive pilot and feedback overhead, the TRS period $ T_{\rm period}^{\rm trs} $ and the feedback period $ T_{\rm period}^{\rm csi\text{-}fb} $ should be designed in accordance with the temporal characteristics of S2G channels. In the following, we analyze the design criteria for pilot and feedback periods.  

\subsubsection{Prediction Model Validity}
Owing to the deterministic orbital motion of satellites, the temporal evolution of S2G channel parameters can be characterized by the prediction model over different time scales, as will be detailed in \secref{sec:channel model}. 
Let $T_{\rm eff}$ denote the validity interval within which the approximation error of the adopted prediction model remains below a prescribed threshold $\epsilon_{d}$. 
To ensure reliable prediction, the CSI acquisition and feedback must be completed within this validity interval, implying:  
\begin{align}
  \textstyle M T_{\rm period}^{\rm trs} + T_{\rm period}^{\rm csi\text{-}fb} < T_{\rm eff}.
\end{align}
This condition reveals the fundamental trade-off between overhead reduction and model fidelity. 

\subsubsection{Phase Unwrapping Effectiveness}
Coherent transmission hinges on phase synchronization across cooperative satellites. 
Since phase observations are inherently cyclic, reliable phase prediction requires PU across successive TRS transmissions. 
To avoid unwrapping failure, the phase rotation over one inter-TRS interval caused by maximum Doppler shift estimation error $ \Delta \nu_{\rm max} $ must remain below $ \pi $, i.e.,  
\begin{align}
  \textstyle \Delta \nu_{\rm max} T_{\rm period}^{\rm trs} < 0.5.
\end{align}
This condition guarantees that the inter-TRS phase evolution remains continuous from the perspective of PU.

\vspace{-2mm}
\section{Multi-Satellite Signal Model}
\label{sec: mSat signal model}

This section establishes a time-varying S2G channel model, constructing a polynomial model to capture the temporal evolution of propagation delays and Doppler shifts. 
Subsequently, we derive an OFDM-based multi-satellite signal model that accounts for non-ideal synchronization, followed by a detailed phase analysis of equivalent TF-domain channels. 

\vspace{-2mm}
\subsection{Channel Model}
\label{sec:channel model}
The baseband channel impulse response (CIR) between the $ s $-th satellite and the $ u $-th UE \cite{xiang2024massive,zhu2024joint} is denoted as  
\begin{align}
  \label{eq:CIR channel}
  \scalebox{0.95}{ $\displaystyle
  \!\!\tilde{\mathbf{h}}_{su}(t,\tau) 
  \hspace{-1mm}=\hspace{-2mm} \sum_{l=0}^{L_{su}-1} \hspace{-2mm}
  \alpha_{sul}(t) 
  e^{\bar{\jmath}\psi_{sul}}
  e^{-\bar{\jmath}2\pi f_{\rm c}\tau_{sul}(t)}  \delta(\tau \!-\! \tau_{sul}(t))
  \mathbf{v}(\bm{\theta}_{sul}(t)),
  $}
\end{align} 
where $ L_{su} $ denotes the number of multipath components. 
$ \psi_{sul} $, $ \alpha_{sul}(t) $, and $ \tau_{sul}(t) $ represent the original phase, real-valued gain, and propagation delay of the $ l $-th path. 
The corresponding Doppler shift at $ f_{\rm c} $ is denoted as $ \nu_{sul}(t) = -\frac{{\rm d}\tau_{sul}(t)}{{\rm d}t} f_{\rm c} $. 
$ \bm{\theta}_{sul}(t) = [\theta_{sul}^{\rm ele}(t),\theta_{sul}^{\rm azi}(t)]^{T} $ denotes the elevation and azimuth angles of departure (AoDs). 
The steering vector is given~by 
\begin{align}
  \mathbf{v}(\bm{\theta}_{sul}(t)) = 
  \mathbf{v}_{N_{\rm x}}\left( \theta_{sul}^{\rm x}(t) \right)  
  \otimes 
  \mathbf{v}_{N_{\rm y}}\left( \theta_{sul}^{\rm y}(t) \right),
\end{align}
with $ \mathbf{v}_{N}(x) = [1,e^{-\bar{\jmath}\pi x},\cdots, e^{-\bar{\jmath}\pi (N-1) x}]^{T} $,  $ \theta_{sul}^{\rm x}(t) = \mathsf{sin}(\theta_{sul}^{\rm ele}(t))\mathsf{cos}(\theta_{sul}^{\rm azi}(t)) $, $ \theta_{sul}^{\rm y}(t) = \mathsf{sin}(\theta_{sul}^{\rm ele}(t))\mathsf{sin}(\theta_{sul}^{\rm azi}(t)) $. 

Since propagation delay and Doppler shift are determined by the S2G propagation distance and its rate of change, we next characterize the temporal evolution of the S2G distance. 
For brevity, the satellite and UE indices $(s, u)$ are omitted.
Since S2G channels are dominated by the LoS component \cite{38.811}, we focus on the temporal evolution of the LoS distance for stationary UEs. 
As illustrated in \cite{ali2002doppler}, the relative angular velocity between a LEO satellite and a UE remains quasi-static over short time scales. This allows for the approximation $ \omega_{\rm se} \approx \omega_{\rm s} - \omega_{\rm e}\mathsf{cos}\theta_{\rm inc} $, where $ \omega_{\rm s} $ and $ \omega_{\rm e} $ denote the angular velocities of the satellite and Earth in the Earth-centered inertial frame. $ \theta_{\rm inc} $ is the inclination of the satellite's orbit. 
We define $ t = 0 $ to coincide with the~UE's maximum elevation angle $ \beta_{\rm max} $ and minimum geocentric angle $ \gamma_{\rm min} $. The geocentric angle traversed by the satellite is $ \eta = \omega_{\rm se}t $, and the S2G distance is 
\begin{align}
  d_{0}^{\rm sat}(t) = \sqrt{ r_{\rm sat} + r_{\rm ue} - 2r_{\rm sat}r_{\rm ue}\mathsf{cos}\gamma_{\rm min}\mathsf{cos}(\omega_{\rm se}t) }, 
\end{align}
where $ r_{\rm sat} $ and $ r_{\rm ue} $ denote the geocentric distances of the satellite and UE. 

\begin{figure}[t]
  \centering
  \includegraphics[trim=0mm 0.5mm 0mm 0.5mm, clip, width=0.8\linewidth]{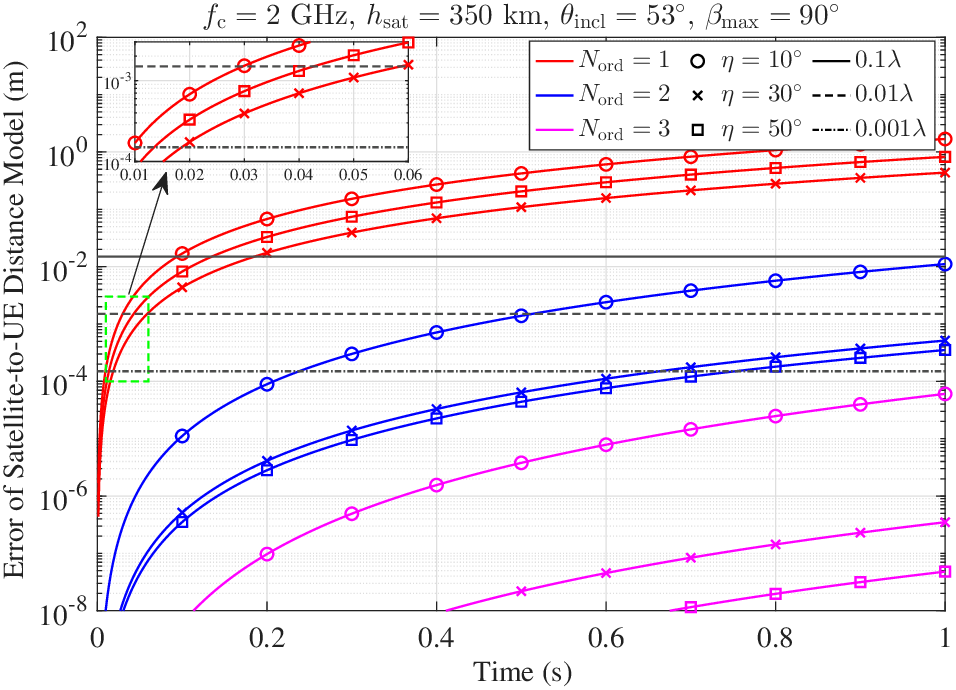}
  \vspace{-3mm}
  \caption{Satellite-to-UE distance modeling error based on Taylor series expansion with $ f_{\rm c} = 2 $ GHz, $ r_{\rm sat} = 6721 $ km, $ \theta_{\rm inc} = 53^{\circ} $, $ \beta_{\rm max} = 90^{\circ} $.}
  \label{fig:SatUeDisApprox}
  \vspace{-1mm}
\end{figure}

To characterize channel evolution over a local observation window, we approximate  $ d_{0}^{\rm sat}(t) $ by an $ N_{\rm ord} $-th-order Taylor series expansion, yielding 
\begin{align}
  \textstyle
  d_{0}^{\rm sat}(t+\Delta t) \approx \sum_{n=0}^{N_{\rm ord}}d_{0}^{{\rm sat},(n)}(t) (\Delta t)^{n} / n!,
\end{align}
where $ d_{0}^{{\rm sat},(n)}(t) $ denotes the $ n $-th-order derivative of $ d_{0}^{\rm sat}(t) $. 
As shown in \figref{fig:SatUeDisApprox}, increasing $ N_{\rm ord} $ extends the valid time scale of the approximation. In particular, for errors below $ 10^{-3}\lambda $, first-, second-, and third-order models are suitable for millisecond-, sub-second-, and second-level intervals.    

\vspace{-1mm}
\begin{remark} 
Under the $N_{\mathrm{ord}}$-th-order approximation of the S2G distance, the propagation delay $\tau^{\mathrm{sat}}_0(t)$ and Doppler shift $\nu^{\mathrm{sat}}_0(t)$ can be represented by time polynomials of order $N_{\mathrm{ord}}$ and $(N_{\mathrm{ord}}-1)$, respectively. This polynomial characterization will serve as the basis for the subsequent signal modeling and channel prediction design. 
\label{remark xxx}
\end{remark}

\subsection{Signal Model}

Based on the design in \secref{sec: framework}, either TRS or coherent transmission calls for TBA- or UE-specific precompensation, implying per-stream OFDM modulation rather than joint modulation of aggregated streams. 
Since the signal models for TRS and coherent transmissions mainly differ in the transmitted signal model, we first develop the TRS transmitted signal model. 
Consider the $ u $-th UE is located in the $ b_{u} $-th BP/TBA. 
Let $ \{ \mathbf{x}_{sb_{u}pk} \}_{k=-N/2}^{N/2-1} $ denote the spatial-frequency-domain symbol vectors from the $ s $-th satellite to the $ b_{u} $-th TBA on the $ p $-th symbol. 
After OFDM modulation, precompensation, and frequency upconversion, the transmitted bandpass signal is 
\begin{align}
  \label{eq:tx_signal_cps}
  &\scalebox{0.95}{ $\displaystyle
  \tilde{\mathbf{x}}_{sb_{u}}^{\rm cps}(t) 
  \!=\! \sum_{p=0}^{P-1} \sum_{k=-N/2}^{N/2-1} 
  \!\! \mathbf{x}_{sb_{u}pk} 
  e^{-\bar{\jmath}\varphi_{sb_{u}pk}^{\rm cps}}
  e^{\bar{\jmath}2\pi (f_{\rm c}+k\Delta f)(t+\tau_{sb_{u}p}^{{\rm cps}}-pT_{\rm sym})}
  $} 
  \nonumber\\  
  &\qquad\qquad
  \scalebox{0.95}{ $\displaystyle
  \cdot e^{-\bar{\jmath}2\pi\nu_{sb_{u}p}^{\rm cps}(t+\tau_{sb_{u}p}^{\rm cps})}
  \mathsf{U}_{\rm tx}(t+\tau_{sb_{u}p}^{{\rm cps}}-pT_{\rm sym}),
  $}
\end{align}
where $ \tau_{sb_{u}p}^{\rm cps} $, $ \nu_{sb_{u}p}^{\rm cps} $, and $ \varphi_{sb_{u}pk}^{\rm cps} $ are the TFP precompensation terms calculated based on the ephemeris and TBA positions. For TRS transmission, $ \varphi_{sb_{u}pk}^{\rm cps} = 0 $. 
$ \mathsf{U}_{\rm tx}(t) $ equals $1$ when $ t\in[-T_{\rm cp},T) $, and $0$ otherwise, accounting for CP insertion. 
The coherent-transmission case $ \tilde{\mathbf{x}}_{su}^{\rm dcps}(t) $ is obtained by replacing these TBA-specific quantities and transmitted symbols with the UE-specific ones, i.e., $ \{ \tau_{sup}^{\rm dcps}, \nu_{sup}^{\rm dcps}, \varphi_{supk}^{\rm dcps} \} $ and $ \mathbf{x}_{supk}^{\rm d} $.

The received bandpass signal is obtained by convolving the transmitted signal \eqref{eq:tx_signal_cps} with the CIR \eqref{eq:CIR channel}, yielding
\begin{align}
  \textstyle
  \tilde{y}_{u}(t) = \sum_{b=0}^{B-1}\sum_{s=0}^{S-1} 
  \int \tilde{\mathbf{h}}_{su}^{T}(t,\tau) 
  \tilde{\mathbf{x}}_{sb}^{\rm cps}(t-\tau) \text{d}\tau. 
\end{align}
After CP removal, the $p$-th-symbol segment is
\begin{align}
  &\tilde{y}_{up}(t) = \tilde{y}_{u}(t+pT_{\rm sym}) \mathsf{U}_{\rm rx}(t),\qquad t\in[0,T),
\end{align}
where $ \mathsf{U}_{\rm rx}(t) $ equals $1$ when $ t\in[0,T) $, and $0$ otherwise. 

The received signal consists of an effective component and an inter-beam interference component. In what follows, we first analyze the effective component. 
Benefitting from the TBA-specific TRS precompensation in \secref{sec:TRS design}, timing and frequency offsets are strictly confined within the CP and the allowable tolerance \cite{38.101-1,moose2002technique,armstrong2002analysis}. 
Accordingly, the effective received baseband signal on the $ p $-th symbol is given by 
\begin{align}
  &
  \scalebox{0.95}{ $\displaystyle
  \tilde{y}_{up}^{\rm eff}(t) = 
  \sum_{s=0}^{S-1} \sum_{l=0}^{L_{su}-1} \sum_{k=-N/2}^{N/2-1}
  \alpha_{supl}(t) 
  e^{\bar{\jmath}\psi_{sul}}
  e^{-\bar{\jmath}2\pi f_{\rm c}(\tau_{supl}(t)-\tau_{sb_{u}p}^{{\rm cps}})}
  $} \nonumber\\
  &
  \scalebox{0.95}{ $\displaystyle
  \cdot e^{\bar{\jmath}2\pi k\Delta f(t-(\tau_{supl}(t)-\tau_{sb_{u}p}^{{\rm cps}}))}
  e^{-\bar{\jmath}2\pi\nu_{sb_{u}p}^{\rm cps}(t+pT_{\rm sym}-(\tau_{supl}(t)-\tau_{sb_{u}p}^{{\rm cps}}))} 
  $}
  \nonumber\\
  &
  \scalebox{0.95}{ $\displaystyle
  \cdot e^{-\bar{\jmath}\varphi_{sb_{u}pk}^{\rm cps}} 
  \mathbf{v}^{T}(\bm{\theta}_{supl}(t)) 
  \mathbf{x}_{sb_{u}pk},\qquad t\in [0,T), 
  $}
\end{align}
where $ \alpha_{supl}(t) = \alpha_{sul}(t+pT_{\rm sym}) $, $ \tau_{supl}(t) = \tau_{sul}(t+pT_{\rm sym}) $, and $ \bm{\theta}_{supl}(t) = \bm{\theta}_{sul}(t+pT_{\rm sym}) $. 

Following the analysis of S2G distance in \secref{sec:channel model}, we assume that the Doppler shift, AoDs, and path gain are quasi-static within one symbol \cite{you2020massive}, i.e., $ \nu_{supl}(t) = \nu_{supl} $, $ \bm{\theta}_{supl}(t) = \bm{\theta}_{supl} $, and $ \alpha_{supl}(t) = \alpha_{supl} $.  
Thus, the propagation delay over one symbol is linearized as $ \tau_{supl}(t) = \tau_{supl} - \nu_{supl}t / f_{\rm c} $, where $ \tau_{supl} $ is the delay at the symbol beginning.  
Since the satellite-to-UE distance is much larger than the scatterer separation \cite{you2020massive}, the AoDs across paths are nearly identical, i.e., $ \bm{\theta}_{supl} \approx \bm{\theta}_{sup0},\forall l $. 
We further define the spatial-frequency-domain symbol vector as $ \mathbf{x}_{sb_{u}pk} = \mathbf{w}_{sb_{u}p}d_{sb_{u}pk} $, where $ \mathbf{w}_{sb_{u}p} $ represents the precoder of the $ s $-th satellite for the $ b_{u} $-th BP. 
We assume that the satellites employ the angle-based precoder using the ephemeris and BP center positions, yielding $ \mathbf{w}_{sb_{u}p} = \sqrt{P_{\rm tx}}\mathbf{v}^{*}(\bm{\theta}_{sb_{u}p}^{\rm b}) $, where $ \bm{\theta}_{sb_{u}p}^{\rm b} $ is the AoD vector from the $ s $-th satellite to the $ b_{u} $-th BP center on the $p$-th symbol. 
Sampling the effective received signal $ \tilde{y}_{up}^{\rm eff}(t) $ yields 
\begin{align}
  & 
  \scalebox{0.95}{ $\displaystyle
  \!\!\tilde{y}_{up}^{\rm eff}[n] = \sum_{s=0}^{S-1} \sum_{l=0}^{L_{su}-1} \sum_{k=-N/2}^{N/2-1}
  \alpha_{supl}^{\rm res}
  e^{\bar{\jmath}\tilde{\psi}_{supl}^{\rm res}} 
  e^{\bar{\jmath}2\pi \frac{n}{N\Delta f}\tilde{\nu}_{supl}^{\rm res}} 
  $}
  \nonumber\\
  & 
  \scalebox{0.95}{ $\displaystyle
  \!\!\cdot 
  e^{-\bar{\jmath}2\pi \frac{k}{NT_{\rm s}}\tilde{\tau}_{supl}^{\rm res}} 
  e^{\bar{\jmath}2\pi\frac{nk}{N}(1+\frac{{\nu}_{supl}}{f_{\rm c}})}
  e^{-\bar{\jmath}\varphi_{sb_{u}pk}^{\rm cps}} d_{sb_{u}pk}, \: n\in \mathcal{N},
  $}
\end{align}
where $ \mathcal{N} = \{ n\in\mathbb{Z} | 0 \le n \le N-1 \} $. 
The residual phase, residual delay, residual Doppler shift, and equivalent real-valued path gain are defined as follows:
\begin{subequations}
  \label{eq:residual param}
  \begin{align}
    \label{eq:phase res}
    &\tilde{\psi}_{supl}^{\rm res} =  \psi_{sul} + \psi_{sb_{u}p}^{\rm b} - \varphi_{sb_{u}pk}^{\rm cps} \\ \nonumber
    &\qquad - 2\pi ( (f_{\rm c}-\nu_{sb_{u}p}^{\rm cps})\tilde{\tau}_{supl}^{\rm res} + pT_{\rm sym}\nu_{sb_{u}p}^{\rm cps} ), \\
    &\tilde{\tau}_{supl}^{\rm res} = \tau_{supl} - \tau_{sb_{u}p}^{\rm cps}, \\
    &\tilde{\nu}_{supl}^{\rm res} = \nu_{supl} - \nu_{sb_{u}p}^{\rm cps}, \\
    &\alpha_{supl}^{\rm res} = \alpha_{supl}\alpha_{sb_{u}p}^{\rm b},
  \end{align} 
\end{subequations}
where $ \psi_{sb_{u}p}^{\rm b} $ and $ \alpha_{sb_{u}p}^{\rm b} $ are the phase and array gain due to precoding, i.e., $ \mathbf{v}^{T}(\bm{\theta}_{sup0}) \mathbf{w}_{sb_{u}p} = \alpha_{sb_{u}p}^{\rm b}e^{\psi_{sb_{u}p}^{\rm b}} $. 
The above analysis is consistent with the results in \cite{wang2025statistical,wang2025MSMS}. 

\begin{figure*}[t]
  \centering
  \subfigure[]{%
  \includegraphics[
      height=0.25\textwidth,
      trim=10 22 10 20,
      clip
  ]{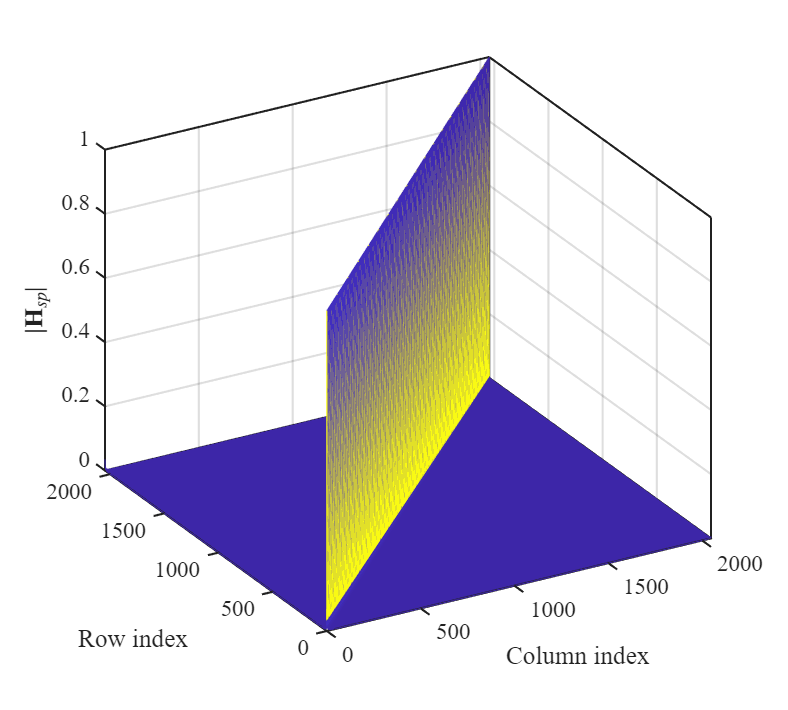}
  \label{fig:sub_a}
  }
  \hspace{4mm}
  \subfigure[]{%
  \includegraphics[
      height=0.25\textwidth,
      trim=0 12 25 5,
      clip
  ]{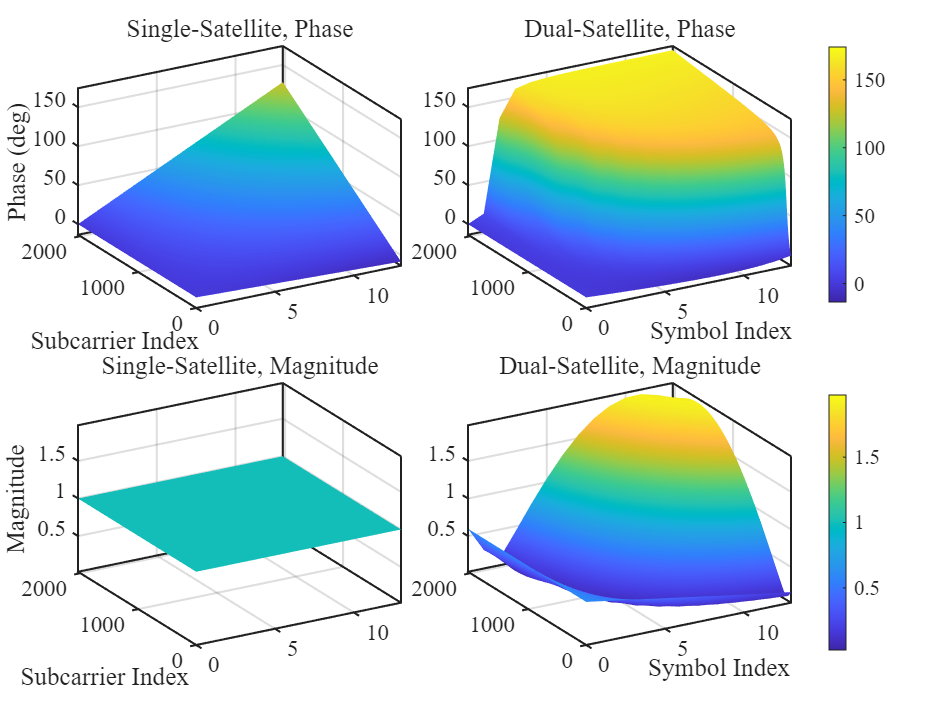}
  \label{fig:sub_b}
  }
  \hspace{4mm}
  \subfigure[]{%
  \includegraphics[
      height=0.25\textwidth,
      trim=2 5 15 7,
      clip
  ]{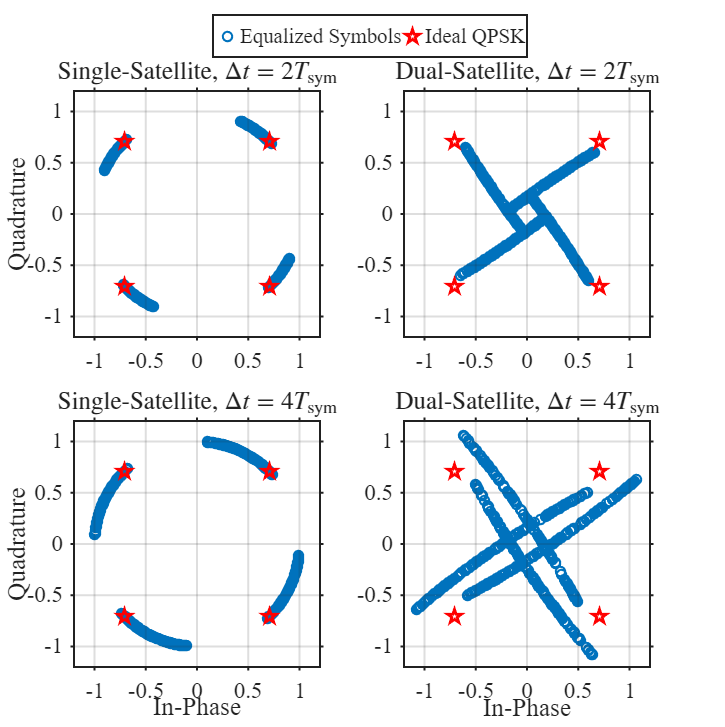}
  \label{fig:sub_c}
  }
  \vspace{-2mm}
  \caption{Characteristic analysis of equivalent TF-domain channels: (a) ICI power of single-satellite channel; (b) phase and magnitude variations of single-satellite channel (left) and dual-satellite composite channel (right); (c) equalization results over single-satellite channel (left) and dual-satellite composite channel (right). Parameters: $f_{\rm c}\!=\!2$ GHz, $N\!=\!2048$, $\Delta f\!=\!15$ kHz, $\nu_{00}\!=\!27$ kHz, $\nu_{10}\!=\!-20$ kHz, $\tilde{\nu}_{00}^{\rm res}\!=\!180$ Hz, $\tilde{\nu}_{10}^{\rm res}\!=\!-260$ Hz, $\alpha_{00}^{\rm res}\!=\!1$, and $\alpha_{10}^{\rm res}\!=\!0.92$.}
  \label{fig:three_subfigs}
  \vspace{-3.5mm}
\end{figure*}

After OFDM demodulation, the effective received signal vector of the $ u $-th UE on the $ p $-th symbol can be expressed as
\begin{align}
  \label{eq:effect signal}
  \mathbf{y}_{up}^{\rm eff} 
  &
  \scalebox{0.95}{ $\displaystyle
  \!=\! \sum_{s=0}^{S-1} 
  \sum_{l=0}^{L_{su}-1}
  \!\!\bm{\Xi} 
  \mathbf{H}_{supl} 
  \mathbf{d}_{sb_{u}p}
  ,
  $}
\end{align}
where $\bm{\Xi}\in\mathbb{B}^{N_{\rm trs}\times N}$ selects the subcarriers in~$\mathcal{N}_{\rm trs}$, i.e., the selected columns form $\mathbf{I}_{N_{\rm trs}}$. 
$ \mathbf{d}_{sb_{u}p}\in\mathbb{C}^{N} $ denotes the symbol vector, with~$ [\mathbf{d}_{sb_{u}p}]_{k}\!=\!d_{sb_{u}pk} $. The equivalent TF-domain channel matrix is given by 
\begin{align}
  \mathbf{H}_{supl} \!=\! 
  \alpha_{supl}^{\rm res}
  e^{\bar{\jmath}\tilde{\psi}_{supl}^{\rm res}}
  \tilde{\mathbf{F}}_{N} 
  \bm{\Lambda}_{\nu}(\tilde{\nu}_{supl}^{\rm res})
  \check{\mathbf{F}}_{N}^{H}(\nu_{supl}) 
  \mathsf{diag}\{ \mathbf{f}_{\tau}(\tilde{\tau}_{supl}^{\rm res}) \},
\end{align}
where  
$ \tilde{\mathbf{F}}_{N} $ is the phase-shift discrete Fourier transform matrix of size $ N $ with $ [\tilde{\mathbf{F}}_{N}]_{k,n} = \frac{1}{\sqrt{N}}e^{-\bar{\jmath}2\pi \frac{(k-N/2)n}{N}} $. 
The diagonal matrix $ \bm{\Lambda}_{\nu}(\tilde{\nu}_{supl}^{\rm res}) $ captures the residual Doppler shift, with $ [\bm{\Lambda}_{\nu}(\tilde{\nu}_{supl}^{\rm res})]_{n,n} = e^{\bar{\jmath}2\pi \frac{n}{N\Delta f}\tilde{\nu}_{supl}^{\rm res}} $. 
$ \check{\mathbf{F}}_{N}(\nu_{supl})\mathsf{diag}\{ \mathbf{f}_{\tau}(\tilde{\tau}_{supl}^{\rm res}) \} $ incorporates the Doppler shift-induced phase rotation, with $ [\check{\mathbf{F}}_{N}(\nu_{supl})]_{k,n} \!=\! \frac{1}{\sqrt{N}}e^{-\bar{\jmath}2\pi \frac{(k-N/2)n}{N}(1+\frac{\nu_{supl}}{f_{\rm c}})} $ and $ [\mathbf{f}_{\tau}(\tilde{\tau}_{supl}^{\rm res})]_{k} \!=\! e^{-\bar{\jmath}2\pi \frac{k-N/2}{NT_{\rm s}}\tilde{\tau}_{supl}^{\rm res}} $. 
The effective received signal model for coherent transmission is obtained analogously by replacing the TBA-specific precompensation and symbol vector with the UE-specific ones, i.e., $ \{ \tau_{sup}^{\rm dcps}, \nu_{sup}^{\rm dcps}, \bm{\varphi}_{sup}^{\rm dcps} \} $ and $ \mathbf{d}_{sup}^{\rm d} $, yielding $ \mathbf{y}_{up}^{\rm deff} $ and $ \mathbf{H}_{supl}^{\rm d} $.  

In addition to the effective component, the received signal contains inter-beam interference, which is given by  
\begin{align}
  \textstyle \mathbf{z}_{up}^{\rm int} = \sum_{b\neq b_{u}}\sum_{s=0}^{S-1}\sum_{l=0}^{L_{su}-1} \mathbf{z}_{supl,b}^{\rm int},
\end{align}
where $ \mathbf{z}_{supl,b}^{\rm int} \in \mathbb{C}^{N_{\rm trs}} $ is the interference of the $ u $-th UE on the $ p $-th symbol caused by the transmission from the $ s $-th satellite to the $ b $-th $ (b\neq b_{u}) $ TBA via the $ l $-th path. 
Owing to the substantial differential propagation delays and Doppler shifts between satellites and UEs, $ \mathbf{z}_{up}^{\rm int} $ can be modeled as Gaussian distribution \cite{geng2015optimality,wang2025statistical} with variance $ \sigma_{{\rm int},up}^{2} \!=\! \mathsf{E}
[ \sum_{b\neq b_{u}}\sum_{s=0}^{S-1}\sum_{l=0}^{L_{su}-1} (\alpha_{supl,b}^{\rm res})^{2} ] $, where $ \alpha_{supl,b}^{\rm res} = \alpha_{supl} \sqrt{P_{\rm tx}} |\mathbf{v}^{T}(\bm{\theta}_{sup0})\mathbf{v}^{*}(\bm{\theta}_{sbp}^{\rm b})|  $. 
The interference power is determined by the residual AoDs and can be effectively suppressed by beam-hopping design \cite{you2020massive}. 

Based on the above analysis, the received signal of the $ u $-th UE on the $ p $-th symbol can be summarized as 
\begin{align}
  \label{eq:rece sig}
  \mathbf{y}_{up} = \mathbf{y}_{up}^{\rm eff} + \mathbf{z}_{up},
\end{align}
where $ \mathbf{z}_{up} = \mathbf{z}_{up}^{\rm nse} + \mathbf{z}_{up}^{\rm int} $ represents the complex Gaussian noise-plus-interference vector, with independent and identically distributed (i.i.d.) elements $ \mathsf{CN}([\mathbf{z}_{up}]_{k};0,\sigma_{\rm nse}^{2}+\sigma_{{\rm int},up}^{2}) $. 
$ \mathbf{z}_{up}^{\rm nse} $ is the Gaussian noise vector with variance $ \sigma_{\rm nse}^{2} $.  

\vspace{-3mm}
\subsection{Phase Analysis of Time-Frequency-Domain Channels} 
\label{sec:phase analysis} 

The performance of multi-satellite coherent transmission critically depends on the phase alignment among signals from different satellites. Since the S2G channels are typically dominated by the LoS component, we focus on the phase characteristics of the equivalent TF-domain channel associated with the LoS path. For notational simplicity, the UE and path indices $(u,l)$ are omitted. To facilitate the analysis, we assume that time-frequency precompensation is performed at the $0$-th symbol and remains fixed over the subsequent symbols. 

We first analyze the impact of the matrix $ \boldsymbol{\Phi}_{sp}^{\nu} = \tilde{\mathbf{F}}_{N} \bm{\Lambda}_{\nu}(\tilde{\nu}_{sp}^{\rm res}) \check{\mathbf{F}}_{N}^{H}(\nu_{sp}) $ on the channel phase. 
The off-diagonal and diagonal elements of $ \boldsymbol{\Phi}_{sp}^{\nu} $ characterize the ICI and the phase shift, respectively. 
Since the Doppler shift is small in the FFT-size-scaled carrier-normalized sense, i.e., $ |\frac{N\nu_{sp}}{2f_{\rm c}}| \ll 1 $, and the TBA/UE-specific precompensation further suppresses the residual Doppler shift such that $ |\frac{\tilde{\nu}_{sp}^{\rm res}}{\Delta f}| \ll 1 $, the magnitudes of the off-diagonal elements of $ \boldsymbol{\Phi}_{sp}^{\nu} $ are negligible compared with those of the diagonal elements, as shown in \figref{fig:sub_a}. 
Furthermore, given the relatively low signal-to-noise ratio (SNR) regime of power-limited SatCom links, the resulting ICI power under the considered system parameters is much lower than the noise floor, and can therefore be neglected. 
Accordingly, $ \boldsymbol{\Phi}_{sp}^{\nu} $ can be approximated as a diagonal matrix, and its diagonal elements can be further approximated by Taylor expansion as
\begin{align}
  \label{eq:approx}
  \textstyle
  [\boldsymbol{\Phi}_{sp}^{\nu}]_{k,k} 
  &\textstyle \approx
  e^{\bar{\jmath}\pi(k-\frac{N}{2})\frac{N-1}{N}\frac{\nu_{sp}}{f_{\rm c}} } \cdot e^{\bar{\jmath}\pi\frac{N-1}{N} \frac{\tilde{\nu}_{sp}^{\rm res}}{\Delta f} }.  
\end{align} 

Accordingly, the equivalent single-satellite channel on the $p$-th OFDM symbol can be compactly expressed as 
\begin{align}
  \label{eq:channel approx}
  \mathbf{H}_{sp}  \approx \mathsf{diag}\{ \mathbf{h}_{sp} \} = \alpha_{sp}^{\rm res} e^{\bar{\jmath}\psi_{sp}^{\rm res}} \mathsf{diag}\{ \mathbf{f}_{\tau}(\tau_{sp}^{\rm res}) \}. 
\end{align}
Here, the equivalent residual delay and phase are defined as
\begin{subequations}
  \begin{align}
    \label{eq:res delay with wideband effect}
    &\textstyle \tau_{sp}^{\rm res} = \tilde{\tau}_{s0}^{\rm res} - \sum_{p^{\prime}=1}^{p} \frac{\nu_{sp^{\prime}}}{f_{\rm c}}T_{\rm sym} - \frac{N-1}{2N\Delta f} \frac{\nu_{sp}}{f_{\rm c}},\\
    \label{eq:res phase with res dop}
    &\textstyle \psi_{sp}^{\rm res} = \tilde{\psi}_{s0}^{\rm res} + 2\pi\sum_{p^{\prime}=1}^{p} \tilde{\nu}_{sp^{\prime}}^{\rm res} T_{\rm sym} + \pi\frac{N-1}{N}\frac{\tilde{\nu}_{sp}^{\rm res}}{\Delta f}, 
  \end{align}
\end{subequations}
which indicate that the Doppler shift mainly induces a frequency-dependent linear phase term, whereas the residual Doppler shift contributes a frequency-independent phase term. Moreover, due to temporal accumulation, both terms evolve with time, and the former further enlarges the phase difference across subcarriers. This expression shows that the single-satellite channel phase evolves in a relatively regular manner over time and frequency, as shown in \figref{fig:sub_b}. 

However, for multi-satellite transmission, 
the phase of the composite channel on the $p$-th symbol and the $k$-th subcarrier becomes 
\begin{align}
  \textstyle 
  \angle( \sum_{s=0}^{S-1} \alpha_{sp}^{\rm res}e^{-\bar{\jmath}2\pi (k-\frac{N}{2})\Delta f \tau_{sp}^{\rm res} } \cdot e^{\bar{\jmath}\psi_{sp}^{\rm res}} ), 
\end{align}
where $ \angle(\cdot) $ is the phase-angle extraction operator. 
Since the signals from different satellites generally experience distinct path gains and Doppler-induced phase evolutions, 
the inter-satellite relative phase mismatch increases jointly over time and frequency. 
Consequently, the composite-channel phase and magnitude no longer exhibit the regular structure of the single-satellite case, but instead become nonlinear TF-varying quantities, as shown in \figref{fig:sub_b}. Such phase misalignment leads to coherent combining loss and degrades the effective signal-to-interference-plus-noise ratio (SINR).

This effect is further illustrated by the equalization results in \figref{fig:sub_c}. 
Assume that the demodulation reference signal (DMRS) is placed on the $0$-th OFDM symbol. The UE obtains the channel estimate on the $0$-th symbol from the DMRS and then performs one-tap equalization for the received signals on the subsequent symbols using the nearest-neighbor hold interpolation method. To isolate the effect of the channel on the equalization results, we assume noise-free pilot-based channel estimation and equalization. 
In the single-satellite case, the equalized symbols mainly undergo a regular phase rotation, which can still be compensated at the receiver. In contrast, in the multi-satellite case, the superposition of multiple asynchronously evolving phases causes irregular amplitude and phase distortions, which become more severe over time $ \Delta t $ and cannot be effectively removed by receiver-side postprocessing. Therefore, accurate coherent transmission requires fine-grained phase precompensation at the transmitter, performed on a symbol- and subcarrier-wise basis. 

\section{Detailed Design of the JCAFPS Framework}
\label{sec: JCAFPS}

This section presents the algorithmic realization of the proposed JCAFPS framework. The design consists of three stages: single-TRS-based channel parameter estimation, multi-TRS-based channel prediction, and predictive CSI feedback with UE-specific TFP precompensation. 

\subsection{Single-TRS-Based Channel Parameter Estimation}
\label{sec:channel estimation} 

From one TRS transmission, the UE estimates the delay, phase, gain, and Doppler shift of each satellite-to-UE link with sufficient accuracy to support subsequent channel prediction. This is achieved through initial channel estimation, super-resolution delay-domain parameter extraction, and coarse Doppler shift estimation. 

\subsubsection{Initial Channel Estimation} 
Combining \eqref{eq:effect signal}, \eqref{eq:rece sig},and \eqref{eq:channel approx}, the demodulated frequency-domain received signal of the $ u $-th UE on the $p$-th symbol of the $m$-th TRS transmission can be approximated as 
\begin{align}
  \label{eq:approx eff sig}
  \textstyle
  \mathbf{y}_{up}^{(m)} \approx
  \sum_{s=0}^{S-1} \mathsf{diag}\{ \mathbf{d}_{sb_{u}p}^{{\rm trs},(m)} \}  
  \mathbf{h}_{sup}^{{\rm trs},(m)} + \mathbf{z}_{up}^{{\rm trs},(m)},  
\end{align} 
where $ \mathbf{d}_{sb_{u}p}^{{\rm trs},(m)} =  
\bm{\Xi}\mathbf{d}_{sb_{u}p}^{(m)} $ and $ \mathbf{z}_{up}^{{\rm trs},(m)} = \bm{\Xi}\mathbf{z}_{up}^{(m)} $ denote the TRS vector and the noise vector of the $ u $-th UE on the $p$-th symbol of the $m$-th TRS transmission.  
The equivalent frequency-domain channel response vector is
\begin{align}
  \label{eq:equiv freq channel}
  \mathbf{h}_{sup}^{{\rm trs},(m)} = \mathbf{F}_{\tau}^{\rm trs}(\bm{\tau}_{sup}^{{\rm res},(m)}) \bm{\Lambda}_{\psi}(\bm{\psi}_{sup}^{{\rm res},(m)}) \bm{\alpha}_{sup}^{{\rm res},(m)}, 
\end{align} 
where $ \bm{\alpha}_{sup}^{{\rm res},(m)}\in\mathbb{R}^{L_{su}} $ represents the real-valued delay-domain channel gain vector. 
$ \mathbf{F}_{\tau}^{\rm trs}(\bm{\tau}_{sup}^{{\rm res},(m)})\in\mathbb{C}^{N_{\rm trs}\times L_{su}} $ is the FT matrix, with its $ l $-th column vector $ \mathbf{f}_{\tau}^{\rm trs}(\tau_{supl}^{{\rm res},(m)}) = \bm{\Xi}\mathbf{f}_{\tau}(\tau_{supl}^{{\rm res},(m)}) $. 
We further define the equivalent CIR as $ \mathbf{c}_{sup}^{(m)} = \bm{\Lambda}_{\psi}(\bm{\psi}_{sup}^{{\rm res},(m)}) \bm{\alpha}_{sup}^{{\rm res},(m)} \in\mathbb{C}^{L_{su}} $.  

To eliminate the impact of basic TRS sequence, the UE first performs least-squares (LS) estimation on $ \mathbf{y}_{up}^{(m)} $ as 
\begin{align}
  \label{eq:LS chan est}
  \mathbf{h}_{up}^{{\rm ls},(m)} &= \mathsf{diag}\{ ( \bar{\mathbf{d}}_{b_{u}p}^{{\rm trs},(m)} )^{*} \} \mathbf{y}_{up}^{(m)} \nonumber\\
  &= \mathbf{F}_{\tau}^{\rm trs}(\bm{\tau}_{up}^{{\rm ls},(m)}) \mathbf{c}_{up}^{(m)} + \mathbf{z}_{up}^{{\rm ls},(m)},
\end{align} 
where 
$ \bm{\tau}_{up}^{{\rm ls},(m)} = \bm{\tau}_{up}^{{\rm res},(m)} + \tilde{\bm{\phi}}_{u} $. 
Here, $ \bm{\tau}_{up}^{{\rm res},(m)} = [(\bm{\tau}_{0up}^{{\rm res},(m)})^{T},\ldots,(\bm{\tau}_{(S-1)up}^{{\rm res},(m)})^{T}]^{T} $ denotes the stacked residual delay vector, and $ \tilde{\bm{\phi}}_{u} = [\frac{\phi_{0}}{N_{\rm sc}\Delta f}\mathbf{1}_{L_{0u}}^{T},\ldots,\frac{\phi_{S-1}}{N_{\rm sc}\Delta f}\mathbf{1}_{L_{(S-1)u}}^{T}]^{T} $ represents the delay-domain offset vector introduced by the APS-TRS. 
$\mathbf{c}_{up}^{(m)} = \bm{\Lambda}_{\psi}(\bm{\psi}_{up}^{{\rm res},(m)}) \bm{\alpha}_{up}^{{\rm res},(m)} $ represents the stacked CIR, where $ \bm{\psi}_{up}^{{\rm res},(m)} = [ (\bm{\psi}_{0up}^{{\rm res},(m)})^{T}, \ldots, (\bm{\psi}_{(S-1)up}^{{\rm res},(m)})^{T} ]^{T} $, and $ \bm{\alpha}_{up}^{{\rm res},(m)} = [(\bm{\alpha}_{0up}^{{\rm res},(m)})^{T},\ldots,(\bm{\alpha}_{(S-1)up}^{{\rm res},(m)})^{T}]^{T} $. The equivalent noise-plus-interference vector $ \mathbf{z}_{up}^{{\rm ls},(m)} = \mathsf{diag}\{ ( \bar{\mathbf{d}}_{b_{u}p}^{{\rm trs},(m)} )^{*} \}\mathbf{z}_{up}^{{\rm trs},(m)} $ still follows the i.i.d. complex Gaussian distribution with $ \mathsf{CN}([\mathbf{z}_{up}^{{\rm ls},(m)}]_{k};0,\sigma_{\rm nse}^{2}+\sigma_{{\rm int},up}^{2}) $. 

\begin{remark}
After removing the impact of basic TRS sequence, the satellite-specific phase shifts $ \{\phi_{s}\}_{s=0}^{S-1} $ evenly distribute the equivalent delay-domain channel responses of different satellites, thereby preventing inter-satellite pilot interference. 
\end{remark}

\subsubsection{Channel Parameter Estimation}To support the subsequent channel prediction over multiple TRS transmissions, the UE should first estimate the channel parameter~vector~$  \bm{\zeta}_{up}^{(m)} = [(\bm{\tau}_{up}^{{\rm res},(m)})^{T},(\bm{\psi}_{up}^{{\rm res},(m)})^{T}, (\bm{\alpha}_{up}^{{\rm res},(m)})^{T}]^{T} $ based on the initial channel estimation $ \mathbf{h}_{up}^{{\rm ls},(m)} $. 
Accordingly, the channel parameter estimation problem is formulated as 
\begin{align}
  \label{eq:chan para est}
  \textstyle 
  \underset{\bm{\zeta}_{up}^{(m)}}{\mathsf{min}}\:
  \big\Vert \mathbf{h}_{up}^{{\rm ls},(m)} - \mathbf{F}_{\tau}^{\rm trs}(\bm{\tau}_{up}^{{\rm ls},(m)})
  \mathbf{c}_{up}^{(m)} \big\Vert _{2}^{2}.
\end{align}
The above optimization problem is a delay-domain parametric estimation task \cite{zhu2022ofdm,wu2021channel}. 
Since the factor matrix $ \mathbf{F}_{\tau}^{\rm trs}(\bm{\tau}_{up}^{{\rm ls},(m)}) $ has a Vandermonde matrix with generators $ \{\{ \phi_{supl}^{{\rm gen},(m)} = e^{-\bar{\jmath}2\pi N_{\rm tc}\Delta f \tau_{supl}^{{\rm ls},(m)}} \}_{l=0}^{L_{su}-1}\}_{s=0}^{S-1} $ and accurate delay extraction is required to suppress leakage-induced amplitude and phase distortion, we adopt the spatial-smoothing estimation of signal parameters via rotational invariance techniques(ESPRIT) algorithm \cite{roy2002esprit} algorithm for super-resolution CIR estimation. 

Considering the uniqueness condition of matrix decomposition, the spatial-smoothing parameters $(K_{\rm ss},L_{\rm ss})$ for the spatial-smoothing ESPRIT algorithm must satisfy \cite{kruskal1977three}: 1) $ K_{\rm ss} + L_{\rm ss} = N_{\rm trs} + 1 $ and 2) $ \mathsf{min} \{ K_{\rm ss}-1, L_{\rm ss} \} \ge \sum_{s=0}^{S-1}L_{su} $. 
With these parameters, spatial smoothing \cite{pan2020enhanced} is applied to the initial channel estimate $ \mathbf{h}_{up}^{{\rm ls},(m)} $ for 
\begin{align}
  \label{eq:spatial smoothed channel}
  \textstyle
  \scalebox{0.95}{ $\displaystyle
  \mathbf{H}_{up}^{{\rm ss},(m)} 
  $}
  &\scalebox{0.95}{ $\displaystyle
  \textstyle = \big[ \bm{\Xi}_{0}^{\rm ss}\mathbf{h}_{up}^{{\rm ls},(m)}, \bm{\Xi}_{1}^{\rm ss}\mathbf{h}_{up}^{{\rm ls},(m)}, \ldots, \bm{\Xi}_{L_{\rm ss}-1}^{\rm ss}\mathbf{h}_{up}^{{\rm ls},(m)} \big] 
  $},
\end{align}
where $ \bm{\Xi}_{l_{\rm ss}}^{\rm ss} = [\mathbf{0}_{K_{\rm ss}\times l_{\rm ss}}, \mathbf{I}_{K_{\rm ss}}, \mathbf{0}_{K_{\rm ss}\times (L_{\rm ss}-1-l_{\rm ss})}] $~denotes the selection matrix for the $ l_{\rm ss} $-th column vector. 

Then, we perform the truncated singular value decomposition (SVD) on the smoothed channel matrix $ \mathbf{H}_{up}^{{\rm ss},(m)} $, yielding
\begin{align}
  \label{eq:ESPRIT start}
  \mathbf{H}_{up}^{{\rm ss},(m)} = \check{\mathbf{U}}_{up}^{(m)} \check{\bm{\Sigma}}_{up}^{(m)} (\check{\mathbf{V}}_{up}^{(m)})^{H},
\end{align}
where $ \check{\mathbf{U}}_{up}^{(m)}\in\mathbb{C}^{K_{\rm ss}\times \bar{L}_{u}} $, $ \check{\bm{\Sigma}}_{up}^{(m)} \in \mathbb{R}^{\bar{L}_{u} \times \bar{L}_{u}} $, and $ \check{\mathbf{V}}_{up}^{(m)}\in\mathbb{C}^{L_{\rm ss}\times \bar{L}_{u}} $ denote the truncated left singular vector matrix, singular value matrix, and right singular vector matrix, respectively. 
Note that the number of singular values $ \bar{L}_{u} = \sum_{s=0}^{S-1}\bar{L}_{su} $ is determined by the rank of the channel matrix, which can be estimated by the minimum description length criterion \cite{grunwald2005advances}. 
Given the dominant LoS propagation characteristic of S2G channels \cite{38.811}, the estimation of $ \bar{L}_{u} $ typically equals the number of cooperative satellites. 

Exploiting the Vandermonde structure of $ \mathbf{F}_{\tau}^{\rm trs}(\bm{\tau}_{up}^{{\rm ls},(m)}) $, the standard spatial-smoothing ESPRIT principle implies that the signal subspace satisfies a shift-invariance relation. Accordingly, there exists a unique nonsingular diagonal matrix $ \check{\bm{\Phi}}_{up}^{(m)} = \mathsf{diag}\{ \check{\bm{\phi}}_{up}^{(m)} \} \in\mathbb{C}^{\bar{L}_{u}\times \bar{L}_{u}} $ such that 
\begin{align}
  \label{eq:ESPRIT end}
  \scalebox{0.95}{ $\displaystyle
  \hspace{-2mm}
  \check{\mathbf{M}}_{up}^{(m)} \check{\bm{\Phi}}_{up}^{(m)} (\check{\mathbf{M}}_{up}^{(m)})^{-1} 
  \hspace{-1mm}=\hspace{-1mm}
   ([\check{\mathbf{U}}_{up}^{(m)}]_{0:K_{\rm ss}-2,:})^{\dagger} [\check{\mathbf{U}}_{up}^{(m)}]_{1:K_{\rm ss}-1,:}.
  $}
\end{align}
Eigenvalue decomposition of \eqref{eq:ESPRIT end} yields the normalized eigenvalues, i.e., the estimation of generators $ \hat{\bm{\phi}}_{up}^{{\rm gen},(m)} $, from which the delays are estimated as 
\begin{align}
  \label{eq:delay est1}
  \textstyle
  \hat{\tau}_{supl}^{{\rm ls},(m)} = 
  - \frac{1}{2\pi N_{\rm tc}\Delta f} \angle \hat{\phi}_{supl}^{{\rm gen},(m)}.  
\end{align}
Based on the estimated delays, the factor matrix can be reconstructed, and the CIR can be estimated via LS criterion~as 
\begin{align}
  \label{eq:channel est1}
  \textstyle \hat{\mathbf{c}}_{up}^{(m)} = 
  \big( \mathbf{F}_{\tau}^{\rm trs}(\hat{\bm{\tau}}_{up}^{{\rm ls},(m)}) \big)^{\dagger} \mathbf{h}_{up}^{{\rm ls},(m)}. 
\end{align} 

Since the applied precompensation may vary across symbols, abrupt changes can be introduced into the residual delays and Doppler shifts between adjacent symbols. 
Therefore, the contributions induced by adjustment terms must be removed from the delay and phase expressions in \eqref{eq:delay est1} and \eqref{eq:channel est1}, such that the resulting estimates exhibit temporal dynamics. 
We assume that both satellites and UEs have the knowledge of the ephemeris, BP center positions, and TBA center positions~\cite{xiang2024massive,chen2021system}.
According to \eqref{eq:residual param}, the unadjusted delay, phase, and real-valued gain can be expressed as 
\begin{subequations}
  \label{eq:chan para est2}
  \begin{align}
    \label{eq:delay est2}
    \hat{\tau}_{supl}^{(m)} 
    &= 
    \hat{\tau}_{supl}^{{\rm ls},(m)} 
    - \phi_{s}/(N_{\rm sc}\Delta f)
    + \tau_{sb_{u}p}^{{\rm cps},(m)}, \\
    \label{eq:phase est2}
    \hat{\psi}_{supl}^{{\rm w},(m)} 
    &= 
    \mathsf{unwrap}\{\angle (\hat{c}_{supl}^{(m)} e^{\bar{\jmath}\psi_{sb_{u}p}^{{\rm adj},(m)}} ) \},\\
    \label{eq:gain est2}
    \hat{\alpha}_{supl}^{(m)} &= |\hat{c}_{supl}^{(m)}|/\alpha_{sb_{u}p}^{{\rm b},(m)}, 
  \end{align}
\end{subequations} 
where 
$ \psi_{sb_{u}p}^{{\rm adj},(m)} = -\psi_{sb_{u}p}^{{\rm b},(m)} + 2\pi \big( (f_{\rm c}-\nu_{sb_{u}p}^{{\rm cps},(m)})\tau_{sb_{u}p}^{{\rm cps},(m)} + (mT_{\rm period}^{\rm trs}+pT_{\rm sym} + \frac{N-1}{2N\Delta f})\nu_{sb_{u}p}^{{\rm cps},(m)} \big) $ denotes the phase adjustment. 
$ \mathsf{unwrap}\{ \cdot \} $ represents the PU operation, which guarantees the phase continuity across $ N_{\rm trs}^{\rm sym} $ TRS symbols. 
Since the reciprocal of the TRS symbol spacing in one TRS typically exceeds twice the frequency offset \cite{38.214}, the Nyquist sampling condition is fulfilled, which eliminates phase ambiguity. 
In other words, the PU in \eqref{eq:phase est2} can be executed without any prior knowledge of the frequency offset. 

\subsubsection{Doppler Shift Coarse Estimation}
Based on the phase estimates after intra-TRS PU, we next obtain a coarse Doppler estimate. Over the short duration of one TRS, the propagation delay can be well approximated as a linear function of time according to the validity analysis in \secref{sec:channel model}. Hence, the Doppler shift is treated as constant within one TRS and estimated from the phase rotation across TRS symbols. Neglecting noise, the unadjusted phase is expressed as 
\begin{align}
  \hat{\bm{\psi}}_{sul}^{{\rm w},(m)} = \mathbf{M}^{\psi} [\psi_{sul}^{{\rm ro},(m)}, 2\pi \nu_{sul}^{{\rm ro},(m)}]^{T},
\end{align}
where $ \nu_{sul}^{{\rm ro},(m)} $ and $ \psi_{sul}^{{\rm ro},(m)} $ represent the common Doppler shift and phase. 
The measurement matrix $ \mathbf{M}^{\psi}\in\mathbb{C}^{N_{\rm sym}^{\rm trs}\times 2} $ is defined with its $ (i,j) $-th element $ ( (p_{i}^{\rm trs}-p_{0}^{\rm trs})T_{\rm sym} )^{j} $. 
Therefore, the unadjusted common Doppler shift is estimated~by 
\begin{align}
  \label{eq:residual freq est1}
  \hat{\nu}_{sul}^{{\rm ro},(m)} = [ (\mathbf{M}^{\psi})^{\dagger}(\hat{\bm{\psi}}_{sul}^{{\rm w},(m)}) ]_{1} / 2\pi. 
\end{align}

\subsection{Multi-TRS-based Channel Prediction}
\label{sec:channel prediction}

Based on the channel parameters estimated from individual TRS transmissions, the UE further exploits their temporal correlation across multiple observations to establish a predictive channel model. The objective is to convert discrete historical estimates into a compact continuous-time representation for subsequent feedback and transmitter-side precompensation. 

\subsubsection{Cross-TRS Phase Unwrapping}

Although the PU in \eqref{eq:phase est2} ensures phase continuity within one TRS, it does not resolve the ambiguity across adjacent TRS transmissions, where the inter-TRS interval is much longer. Therefore, cross-TRS PU is required. Specifically, a Doppler estimation error $ \Delta \nu $ causes an inter-TRS phase mismatch $ 2\pi \Delta \nu T_{\rm period}^{\rm trs} $. Hence reliable unwrapping requires $ \Delta \nu T_{\rm period}^{\rm trs} < 0.5 $, which is consistent with the design criterion in \secref{sec:trs design}. 

To improve robustness, we first smooth the coarse Doppler estimates over multiple TRS transmissions through polynomial fitting and then perform cross-TRS PU. Since the cumulative observation interval lies in the sub-second regime, a second-order delay model is sufficient according to the validity analysis in \secref{sec:channel model}. 
Accordingly, neglecting noise, the Doppler shift vector $ \hat{\bm{\nu}}_{sul}^{\rm ro}\in\mathbb{R}^{M} $ follows an $(N_{\rm ord}-1)$-th-order polynomial with coefficient vector $ \mathbf{p}_{sul}^{{\rm ro},\nu}\in\mathbb{R}^{N_{\rm ord}} $, which is estimated by 
\begin{align}
  \label{eq:residual doppler fit1}
  \hat{\mathbf{p}}_{sul}^{{\rm ro},\nu} = (\mathbf{M}^{\nu})^{\dagger} \hat{\bm{\nu}}_{sul}^{\rm ro},
\end{align}
where $ \mathbf{M}^{\nu}\in\mathbb{R}^{M\times N_{\rm ord}} $ denotes the corresponding measurement matrix, with its $ (m,n) $-th element $ (mT_{\rm period}^{\rm trs})^{n} $. 

Using the above estimated polynomial coefficients, cross-TRS PU is performed on the wrapped phases $ \{\{ \hat{\psi}_{supl}^{{\rm w},(m)} \}_{p\in\mathcal{N}_{\rm sym}^{\rm trs}}\}_{m=0}^{M-1} $ to recover their temporal evolution $ \{\{ \hat{\psi}_{supl}^{{\rm uw},(m)} \}_{p\in\mathcal{N}_{\rm sym}^{\rm trs}}\}_{m=0}^{M-1} $. 
Their relationship is given by 
\begin{align}
  \hspace{-3mm}\scalebox{0.95}{ $\displaystyle 
  \hat{\psi}_{sup_{i}^{\rm trs}l}^{{\rm w},(m+1)} \!+\! 2\pi k_{sup_{i}^{\rm trs}l,p}^{(m+1)} 
  \!=\! \hat{\psi}_{supl}^{{\rm uw},(m)} \hspace{-1mm}+\hspace{-1mm}
  \int_{\Delta t_{p}^{(m)}}^{\Delta t_{p_{i}^{\rm trs}}^{(m+1)}}
  \hspace{-3mm} \mathbf{t}_{N_{\rm ord}-1}^{T}
  \hat{\mathbf{p}}_{sul}^{{\rm ro},\nu}
  \text{d}t,
  $}
\end{align}
where $ \mathbf{t}_{N_{\rm ord}-1} = [1,t,\ldots,t^{N_{\rm ord}-1}]^{T} $. 
$ \Delta t_{p_{i}^{\rm trs}}^{(m)} = (p_{i}^{\rm trs}-p_{0}^{\rm trs})T_{\rm sym}+mT_{\rm period}^{\rm trs} $ represents the interval between the $ p_{i}^{\rm trs} $-th symbol of the $ m $-th TRS transmission and the $ p_{0}^{\rm trs} $-th symbol of the $ 0 $-th TRS transmission.  
Note that the $ 0 $-th TRS transmission is used as the reference, such that $ \hat{\psi}_{sup_{i}^{\rm trs}l}^{{\rm uw},(0)} = \hat{\psi}_{sup_{i}^{\rm trs}l}^{{\rm w},(0)} $. 
Due to noise and estimation error, the computed value of the integer variable $ k_{sup_{i}^{\rm trs}l,p_{j}^{\rm trs}}^{(m+1)} $, accounting for phase wrapping, generally deviates from an exact integer.


To refine the estimation, the unwrapped phases $ \{ \hat{\psi}_{supl}^{{\rm uw},(m)} \}_{p\in\mathcal{N}_{\rm sym}^{\rm trs}} $ are employed to derive the integers $ \{ k_{sup_{i}^{\rm trs}l,p}^{(m+1)} \}_{p\in\mathcal{N}_{\rm sym}^{\rm trs}} $, which are ideally identical.  
Practically, we average these $ N_{\rm sym}^{\rm trs} $ independent estimates and apply integer quantization to obtain the final estimate: 
\begin{align}
  \label{eq:integer est}
  \textstyle
  \hat{k}_{sup_{i}^{\rm trs}l}^{(m+1)}  = \mathsf{round}\big\{ \sum_{p\in\mathcal{N}_{\rm sym}^{\rm trs}}k_{sup_{i}^{\rm trs}l,p}^{(m+1)} / N_{\rm sym}^{\rm trs} \big\},
\end{align}
where $ \mathsf{round}\{\cdot\} $ denotes the nearest integer rounding operator. 
Therefore, the unwrapped phase for the $ p_{i}^{\rm trs} $-th symbol of the $ m $-th TRS transmission can be recovered as
\begin{align}
  \label{eq:phase unwrapping}
  \hat{\psi}_{sup_{i}^{\rm trs}l}^{{\rm uw},(m)} 
  = \hat{\psi}_{sup_{i}^{\rm trs}l}^{{\rm w},(m)} + 
  2\pi \hat{k}_{sup_{i}^{\rm trs}l}^{(m)}. 
\end{align}

\subsubsection{Polynomial Coefficient Estimation}
Based on the unwrapped phase and the estimated channel parameters, we next determine the polynomial coefficients that characterize the temporal evolution of phase, delay, and path gain, thereby enabling compact predictive CSI representation. 
Based on \eqref{eq:res phase with res dop}, the unwrapped phase can be characterized by the Doppler shift as  
\begin{align}
  \label{eq:unwrap phase model 1}
  \hspace{-3mm}\scalebox{0.95}{ $\displaystyle 
  \hat{\psi}_{sup_{i}^{\rm trs}l}^{{\rm uw},(m)} \!=\! 
  \psi_{sul}^{\rm ori} \!+\! 
  2\pi \!\!\int_{0}^{\Delta t_{p_{i}^{\rm trs}}^{(m)}}
  \hspace{-2mm}\nu_{sul}(t) \text{d}t   
  \!+\! \pi\frac{N-1}{N\Delta f} \nu_{sul} ( \Delta t_{p_{i}^{\rm trs}}^{(m)} ),
  $}
\end{align}
where $ \psi_{sul}^{\rm ori} $ denotes the original phase at the reference time. $ \nu_{sul}(t) $ is the Doppler shift at time $ t $, which is modeled as an ($N_{\rm ord}-1$)-th-order polynomial,~i.e., 
\begin{align}
  \label{eq:res doppler model}
  \nu_{sul}(t) = 
  \mathbf{t}_{N_{\rm ord}-1}^{T} \mathbf{p}_{sul}^{\nu}, 
\end{align}
where $ \mathbf{p}_{sul}^{\nu}\in\mathbb{R}^{N_{\rm ord}} $ denotes the corresponding polynomial coefficient vector. By combining \eqref{eq:unwrap phase model 1} and \eqref{eq:res doppler model}, the unwrapped phase vector $ \hat{\bm{\psi}}_{sul}^{{\rm uw}}\in\mathbb{R}^{N_{\rm sym}^{\rm trs}M} $ over $ M $ TRS transmissions is expressed by an $ N_{\rm ord} $-th-order polynomial, and the corresponding polynomial coefficient vector $ \mathbf{p}_{sul}^{\psi} = [\psi_{sul}^{\rm ori}, (\mathbf{p}_{sul}^{\nu})^{T}]^{T}\in\mathbb{R}^{N_{\rm ord}+1} $ is estimated by the LS~criterion:  
\begin{align}
  \label{eq:residual phase fit}
  \hat{\mathbf{p}}_{sul}^{\psi} = 
  ( \tilde{\mathbf{M}}^{\psi} )^{\dagger}
  \hat{\bm{\psi}}_{sul}^{{\rm uw}},
\end{align}
where $ \tilde{\mathbf{M}}^{\psi} = [\mathbf{1}_{N_{\rm sym}^{\rm trs}M},\bar{\mathbf{M}}^{\psi}]\in\mathbb{R}^{N_{\rm sym}^{\rm trs}M\times (N_{\rm ord}+1)} $ denotes the corresponding measurement matrix. The $ (i + jN_{\rm sym}^{\rm trs},k) $-th element of $ \bar{\mathbf{M}}^{\psi} $ is defined as $ (\Delta t_{p_{i}^{\rm trs}}^{(j)})^{k}(\frac{2\pi}{k+1}\Delta t_{p_{i}^{\rm trs}}^{(j)} + \pi\frac{N-1}{N\Delta f}) $. 

Following the same polynomial parameterization in \eqref{eq:res delay with wideband effect}, the delay is also determined by the Doppler evolution, which is expressed as 
\begin{align}
  \label{eq:res delay model}
  \hspace{-3mm}\scalebox{0.95}{ $\displaystyle 
  \hat{\tau}_{sup_{i}^{\rm trs}l}^{(m)} \!=\! \tau_{sul}^{\rm ori} \!-\! 
  \frac{1}{f_{\rm c}} \int_{0}^{\Delta t_{p_{i}^{\rm trs}}^{(m)}}
  \hspace{-2mm}\nu_{sul}(t) \text{d}t 
  \!-\! \frac{N-1}{2 f_{\rm c}} 
  T_{\rm s}
  \nu_{sul}(\Delta t_{p_{i}^{\rm trs}}^{(m)}), 
  $}
\end{align} 
where $ \tau_{sul}^{\rm ori} $ denotes the original delay at the reference time instant. 
The delay vector $ \hat{\bm{\tau}}_{sul}\in\mathbb{R}^{N_{\rm sym}^{\rm trs}M} $ over $ M $ TRS transmissions corresponds to an $ N_{\rm ord} $-th-order polynomial with coefficients $ \mathbf{p}_{sul}^{\tau} = [\tau_{sul}^{\rm ori}, (\mathbf{p}_{sul}^{\nu})^{T}]^{T} $, 
whose LS estimate~is  
\begin{align}
  \label{eq:res delay fit}
  \hat{\mathbf{p}}_{sul}^{\tau} = 
  ( \tilde{\mathbf{M}}^{\tau} )^{\dagger} 
  \hat{\bm{\tau}}_{sul}, 
\end{align} 
where the measurement matrix $ \tilde{\mathbf{M}}^{\tau} = [\mathbf{1}_{N_{\rm sym}^{\rm trs}M},\bar{\mathbf{M}}^{\tau}]\in\mathbb{R}^{N_{\rm sym}^{\rm trs}M\times (N_{\rm ord}+1)} $.  
The $ (i + jN_{\rm sym}^{\rm trs},k) $-th element of $ \bar{\mathbf{M}}^{\tau} $ defined as $ -\frac{1}{f_{\rm c}}(\Delta t_{p_{i}^{\rm trs}}^{(j)})^{k}(\frac{1}{k+1}\Delta t_{p_{i}^{\rm trs}}^{(j)} + \frac{N-1}{2}T_{\rm s}) $. 
Given the reduced phase estimation error afforded by PU, $ \hat{\mathbf{p}}_{sul}^{\psi} $ typically exhibits markedly higher accuracy than $ \hat{\mathbf{p}}_{sul}^{\tau} $. 
Thus, we reuse the high-precision coefficients $ [\hat{\mathbf{p}}_{sul}^{\psi}]_{1:N_{\rm ord}} $, and restrict the delay estimation to determining $ \tau_{sul}^{\rm ori} $ with lower complexity~as  
\begin{align}
  \hat{\tau}_{sul}^{\rm ori} \hspace{-1mm}=\hspace{-1mm}
  ([\tilde{\mathbf{M}}^{\tau}]_{:,0})^{\dagger}( \hat{\bm{\tau}}_{sul} \hspace{-1mm}-\hspace{-1mm} [\tilde{\mathbf{M}}^{\tau}]_{:,1:N_{\rm ord}} [\hat{\mathbf{p}}_{sul}^{\psi}]_{1:N_{\rm ord}} ).
\end{align}

Likewise, regarding the temporal evolution of path gain on sub-second scales, we predominantly consider gain fluctuation caused by the dynamic S2G distance, while treating remaining perturbations as noise. 
Thus, the reciprocal of path gain $ \hat{\beta}_{sup_{i}^{\rm trs}l}^{(m)} \!=\! 1/\hat{\alpha}_{sup_{i}^{\rm trs}l}^{(m)} $ is also characterized by Doppler shift~as 
\begin{align}
  \label{eq:gain model}
  \scalebox{0.95}{ $\displaystyle
  \hat{\beta}_{sup_{i}^{\rm trs}l}^{(m)} = 
  \beta_{sul}^{\rm ori} + 
  c_{sul}^{\beta}
  \int_{0}^{\Delta t_{p_{i}^{\rm trs}}^{(m)}}
  \nu_{sul}(t) \text{d}t,
  $}
\end{align}
where the constant $ c_{sul}^{\beta} $ comprises the effect of the carrier frequency, antenna gain, and other factors. $ \beta_{sul}^{\rm ori} $ represents the original reciprocal of path gain at the reference time. 
Therefore, the reciprocal of path gain can be modeled as an $N_{\rm ord}$-th-order polynomial. The ensemble of $M$ TRS-derived gain observations facilitates the estimation of the polynomial coefficient vector $ \mathbf{p}_{sul}^{\beta} = [\beta_{sul}^{\rm ori}, ( c_{sul}^{\beta} \mathbf{p}_{sul}^{\nu} )^{T} ]^{T} $ as
\begin{align}
  \label{eq:gain fit}
  \hat{\mathbf{p}}_{sul}^{\beta} = 
  ( \tilde{\mathbf{M}}^{\beta} )^{\dagger} 
  \hat{\bm{\beta}}_{sul},
\end{align}
where $ \tilde{\mathbf{M}}^{\beta} \in\mathbb{R}^{N_{\rm sym}^{\rm trs}M\times (N_{\rm ord}+1)} $, with the $ (i + jN_{\rm sym}^{\rm trs},k) $-th element given by $ (\Delta t_{p_{i}^{\rm trs}}^{(j)})^{k+1}/(k+1) $. Similarly, we can restrict the path gain estimation to determine $ [\beta_{sul}^{\rm ori}, c_{sul}^{\beta}]^{T} $ by reusing $ [\hat{\mathbf{p}}_{sul}^{\psi}]_{1:N_{\rm ord}} $, i.e., 
\begin{align}
  [\hat{\beta}_{sul}^{\rm ori}, \hat{c}_{sul}^{\beta}]^{T} = 
  \Big( \tilde{\mathbf{M}}^{\beta}
  \Big[
  \begin{array}{cc}
    1 &0 \\
    \mathbf{0}_{N_{\rm ord}} & [\hat{\mathbf{p}}_{sul}^{\psi}]_{1:N_{\rm ord}}
  \end{array}
  \Big]
  \Big)^{\dagger} 
  \hat{\bm{\beta}}_{sul},
\end{align}

\subsection{CSI Feedback and UE-Specific TFP Precompensation} 
\vspace{-1mm}

Based on the polynomial coefficients obtained from channel prediction, the UE periodically feeds back compact predictive CSI $ \{\{ \hat{\mathbf{p}}_{sul}^{\psi}, \hat{\mathbf{p}}_{sul}^{\tau}, \hat{\mathbf{p}}_{sul}^{\beta} \}_{l=0}^{\bar{L}_{su}-1}\}_{s=0}^{S-1} $ to the cooperative satellites. Using these coefficients, satellites can reconstruct the future evolution of the phases $ \hat{\psi}_{supl}^{{\rm uw},(m)} $, Doppler shifts $ \hat{\nu}_{supl}^{(m)} $, delays $ \hat{\tau}_{supl}^{(m)} $, path gains $ \hat{\alpha}_{supl}^{(m)} $, and equivalent frequency-domain channel response vector $ \hat{\mathbf{h}}_{sup}^{{\rm trs},(m)} $ throughout the validity interval via \eqref{eq:unwrap phase model 1}, \eqref{eq:res doppler model}, \eqref{eq:res delay model}, \eqref{eq:gain model}, and \eqref{eq:equiv freq channel}.

This predictive capability empowers the satellites to execute UE-specific TFP precompensation for downlink coherent transmission as follows 
\begin{align}
  \label{eq:TFP precomp}
  \textstyle
  \tau_{sup}^{{\rm dcps},(m)} = \hat{\tau}_{sup0}^{(m)}, \:
  \nu_{sup}^{{\rm dcps},(m)} = \hat{\nu}_{sup0}^{(m)}, \: 
  \varphi_{supk}^{{\rm dcps},(m)} = \angle \hat{h}_{supk}^{{\rm trs},(m)},
\end{align} 
such that the signals from different satellites remain aligned for coherent transmission. In this way, predictive feedback serves as the bridge between UE-side channel prediction and transmitter-side precompensation.
To conclude, the comprehensive JCAFPS framework termed \textbf{ESPRIT-Poly-PU} is summarized in \algref{alg:esprit-poly-pu}. 
 
\begin{figure*}[t]
  \centering
  \subfigure{
    \includegraphics[width=0.18\linewidth]{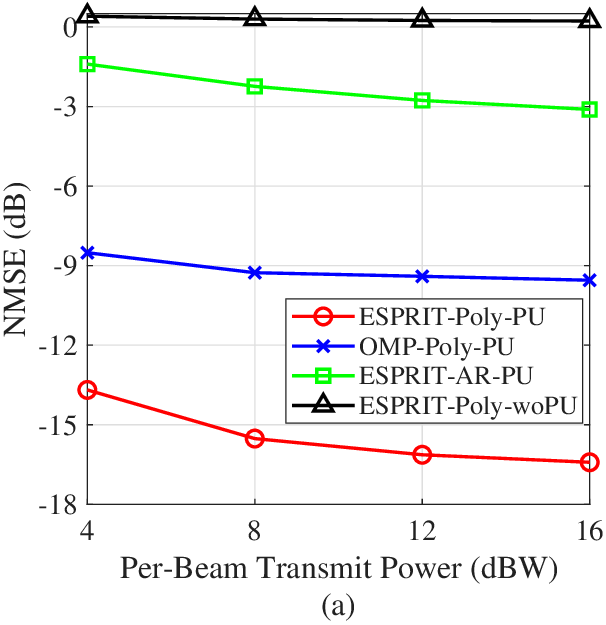}
    \label{fig:NmseVsPt}
  }
  \subfigure{
    \includegraphics[width=0.18\linewidth]{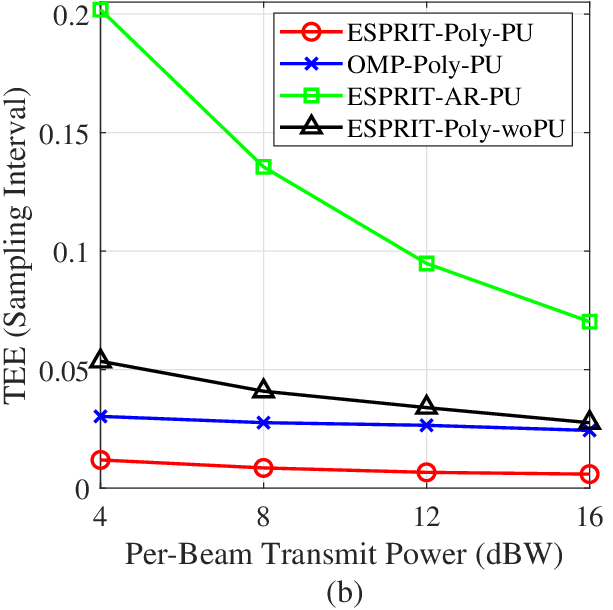}
    \label{fig:TeeVsPt}
  }
  \subfigure{
    \includegraphics[width=0.18\linewidth]{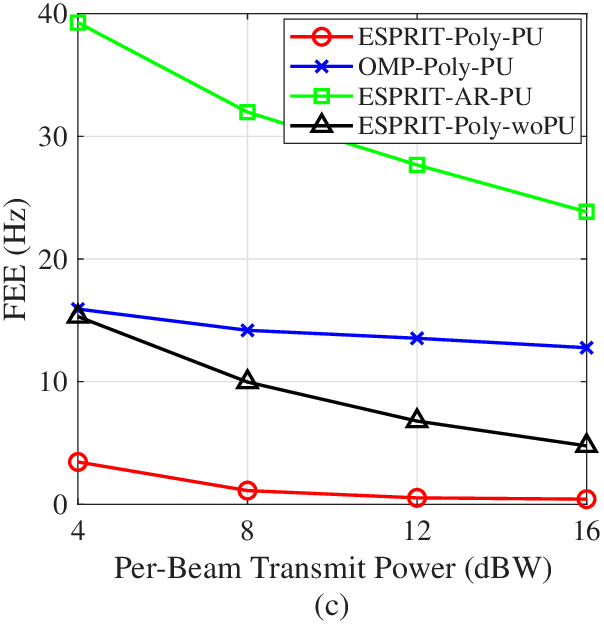}
    \label{fig:FeeVsPt}
  }
  \subfigure{
    \includegraphics[width=0.18\linewidth]{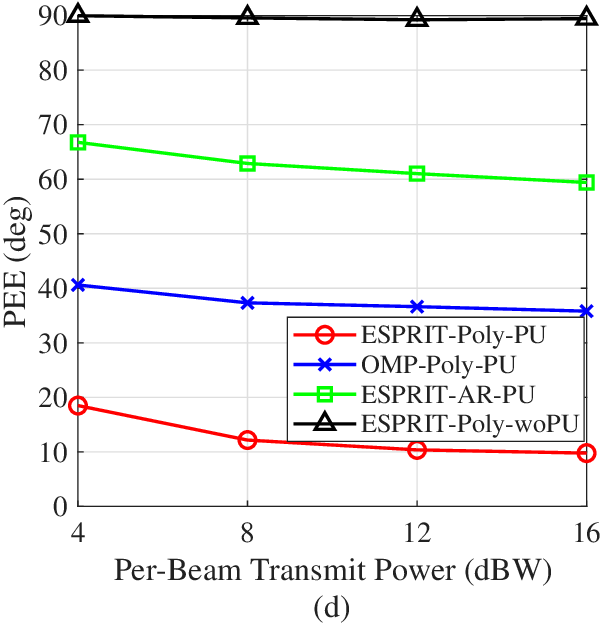}
    \label{fig:PeeVsPt}
  }
  \subfigure{
    \includegraphics[width=0.18\linewidth]{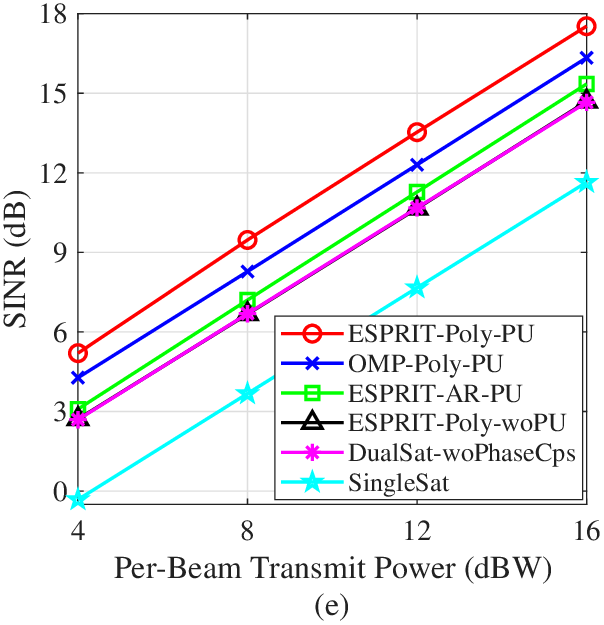}
    \label{fig:SinrVsPt}
  }
  \vspace{-3.5mm}
  \caption{Per-beam power $ P_{\rm beam} $ vs (a) NMSE, (b) TEE, (c) FEE, (d) PEE, and (e) SINR. $ T_{\rm pred} = 80 $ ms, $ M  = 12 $, $ T_{\rm period}^{\rm trs} = 20 $ ms.}
  \label{fig:PerformanceVsPt}
  \vspace{-5mm}
\end{figure*}

\begin{algorithm}[!t]
	\SetAlgoNoLine
	\caption{ESPRIT-Poly-PU Algorithm}
  \label{alg:esprit-poly-pu}
    \textbf{Input:} $\{\{ \mathbf{y}_{up}^{(m)} \}_{p\in\mathcal{N}_{\rm sym}^{\rm trs}}\}_{m=0}^{M-1}$. \\
    \textbf{Output:} $ \{\{ \hat{\mathbf{p}}_{sul}^{\psi}, \hat{\mathbf{p}}_{sul}^{\tau}, \hat{\mathbf{p}}_{sul}^{\beta} \}_{l=0}^{\bar{L}_{su}-1}\}_{s=0}^{S-1} $, 
    $ \{ \tau_{sup}^{{\rm dcps},(m)}, \nu_{sup}^{{\rm dcps},(m)}, \bm{\varphi}_{sup}^{{\rm dcps},(m)} \}_{s=0}^{S-1} $, 
    $ \{ \hat{\mathbf{h}}_{sup}^{{{\rm trs},(m)}} \}_{s=0}^{S-1} $.\\
    \For{$m=0,\cdots M-1$}{
      \textbf{\small\% Single-TRS-Based Channel Parameter Estimation:}\\
      Compute $ \mathbf{h}_{up}^{{\rm ls},(m)} $ via \eqref{eq:LS chan est}, $\forall u,p $\\
      Compute $ \mathbf{H}_{up}^{{\rm ss},(m)} $ via \eqref{eq:spatial smoothed channel}, $ \forall u,p $\\
      Compute the generator $ \hat{\phi}_{supl}^{{\rm gen},(m)} $ via \eqref{eq:ESPRIT start}, \eqref{eq:ESPRIT end}, $ \forall s,u,p,l $\\
			Compute $ \hat{\tau}_{supl}^{(m)}, \hat{\psi}_{supl}^{{\rm w},(m)},\hat{\alpha}_{supl}^{(m)} $ via \eqref{eq:chan para est2}, $\forall s,u,p,l $\\
      Compute $ \hat{\nu}_{sul}^{{\rm ro},(m)} $ via \eqref{eq:residual freq est1}, $\forall s,u,l $\\
    }
    \textbf{\small\% Multi-TRS-Based Channel Prediction:}\\
    Compute $ \hat{\mathbf{p}}_{sul}^{{\rm ro},\nu} $ via \eqref{eq:residual doppler fit1}, $\forall s,u,l $\\
    Compute $\hat{k}_{supl}^{(m)}$, $ \hat{\psi}_{supl}^{{\rm uw},(m)} $ via \eqref{eq:integer est}, \eqref{eq:phase unwrapping}, $\forall s,u,p,l,m $\\ 
    Compute $ \hat{\mathbf{p}}_{sul}^{\psi} $, $ \hat{\mathbf{p}}_{sul}^{\tau} $, $ \hat{\mathbf{p}}_{sul}^{\beta} $ via \eqref{eq:residual phase fit}, \eqref{eq:res delay fit}, \eqref{eq:gain fit}, $\forall s,u,l $\\
    \textbf{\small\% CSI Feedback and UE-Specific TFP Precompensation:}\\
    Feedback $ \hat{\mathbf{p}}_{sul}^{\psi} $, $ \hat{\mathbf{p}}_{sul}^{\tau} $, $ \hat{\mathbf{p}}_{sul}^{\beta} $, $\forall s,u,l $ \\
    Predict $ \hat{\psi}_{supl}^{{\rm uw},(m)} $, $ \hat{\nu}_{supl}^{(m)} $, $ \hat{\tau}_{supl}^{(m)} $, $ \hat{\alpha}_{supl}^{(m)} $ via \eqref{eq:unwrap phase model 1}, \eqref{eq:res doppler model}, \eqref{eq:res delay model}, \eqref{eq:gain model}, $\forall s,u,p,l $ \\
    Predict $ \hat{\mathbf{h}}_{sup}^{{\rm trs},(m)} $ via \eqref{eq:equiv freq channel}, $\forall s,u,p $ \\
    TFP precompensation for coherent transmission: $ \tau_{sup}^{{\rm dcps},(m)}, \nu_{sup}^{{\rm dcps},(m)}, \varphi_{supk}^{{\rm dcps},(m)} $ via \eqref{eq:TFP precomp}, $\forall s,u,p,k $ \\
\end{algorithm} 

\begin{remark}    

Reporting full set~$ \{\{ \hat{\mathbf{p}}_{sul}^{\psi}, \hat{\mathbf{p}}_{sul}^{\tau}, \hat{\mathbf{p}}_{sul}^{\beta} \}_{l=0}^{\bar{L}_{su}-1}\}_{s=0}^{S-1} $ would incur a feedback load of $ 3(N_{\rm ord}\!+\!1)\sum_{s=0}^{S-1}\bar{L}_{su} $~coefficients. Since these parameters' evolution share the same Doppler-related polynomial coefficients, the UE need only report compressed set $ \{\{ \hat{\mathbf{p}}_{sul}^{\psi}, \hat{\tau}_{sul}^{\rm ori}, \hat{\beta}_{sul}^{\rm ori}, \hat{c}_{sul}^{\beta}  \}_{l=0}^{\bar{L}_{su}-1}\}_{s=0}^{S-1} $, reducing the overhead to $ (N_{\rm ord}+4)\sum_{s=0}^{S-1}\bar{L}_{su} $ coefficients while preserving the information needed for TFP-Sync in MSCT. 
\end{remark} 

\section{Simulation Results}
\label{sec:simulation results}

In this section, we present simulation results to illustrate the performance of the proposed framework. 
The scenario and channel parameters are generated using the QuaDRiGa channel simulator, which implements a 3-D geometry-based stochastic channel model \cite{burkhardt2014quadriga,jaeckel20225g,38.811,38.901}. 
Recognized by 3rd generation partnership project (3GPP) as a feature-complete simulator,  QuaDRiGa's satellite-related modeling complies with 3GPP-NTN documents and has been validated via various field tests \cite{burkhardt2014quadriga,jaeckel20225g}. 
Unless otherwise specified, the default system parameters of simulations are listed in Table \ref{tab:system parameters}. 
In each Monte Carlo trial, a reference point is randomly generated within the constellation's service area. The $S$ satellites nearest to this reference point constitute the cooperative cluster. UEs are randomly distributed within the overlapping coverage area of these satellites. 
Regarding the beam hopping strategy, the pattern is determined by selecting $B$ BPs subject to a steering vector quasi-orthogonality constraint \cite{you2020massive}. Besides, the BP radius is defined based on the $3$ dB beam footprint at the~nadir. 

\begin{figure*}[t]
  \centering
  \subfigure{
    \includegraphics[width=0.18\linewidth]{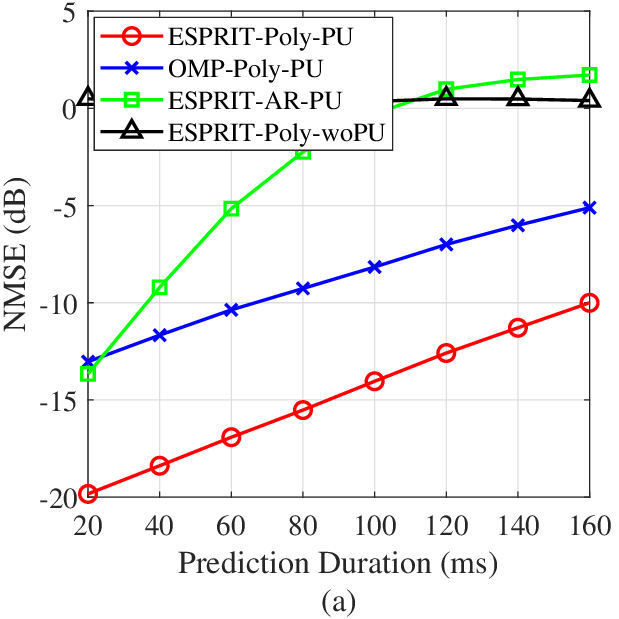}
    \label{fig:NmseVsTime}
  }
  \subfigure{
    \includegraphics[width=0.18\linewidth]{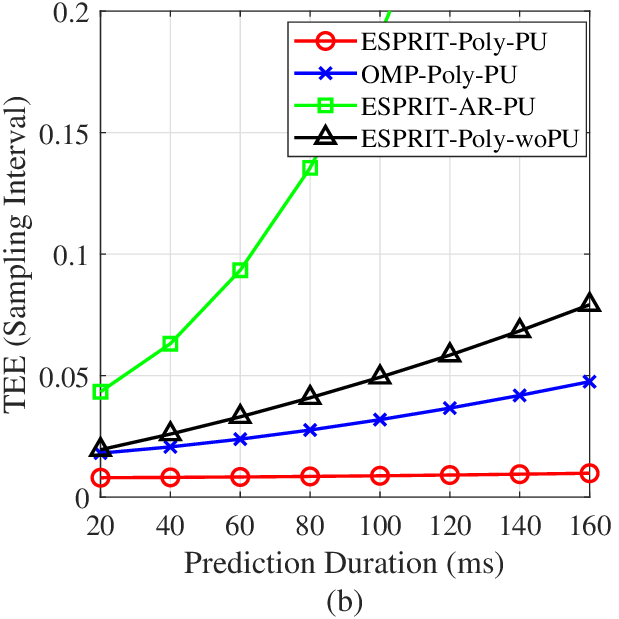}
    \label{fig:TeeVsTime}
  }
  \subfigure{
    \includegraphics[width=0.18\linewidth]{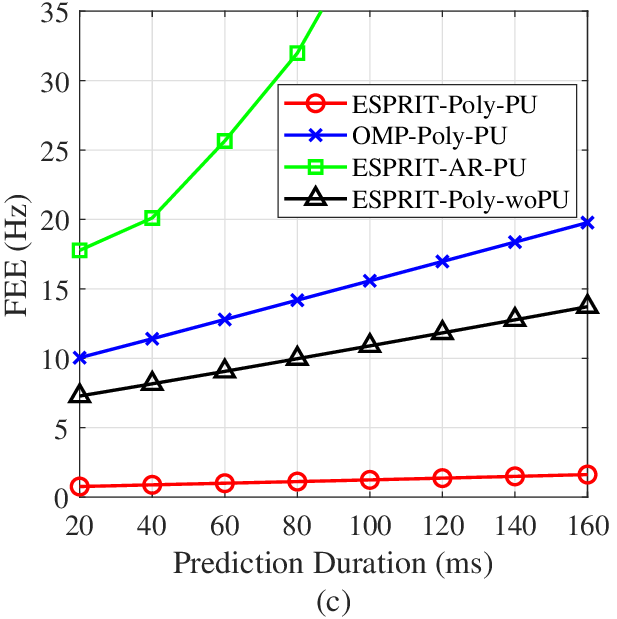}
    \label{fig:FeeVsTime}
  }
  \subfigure{
    \includegraphics[width=0.18\linewidth]{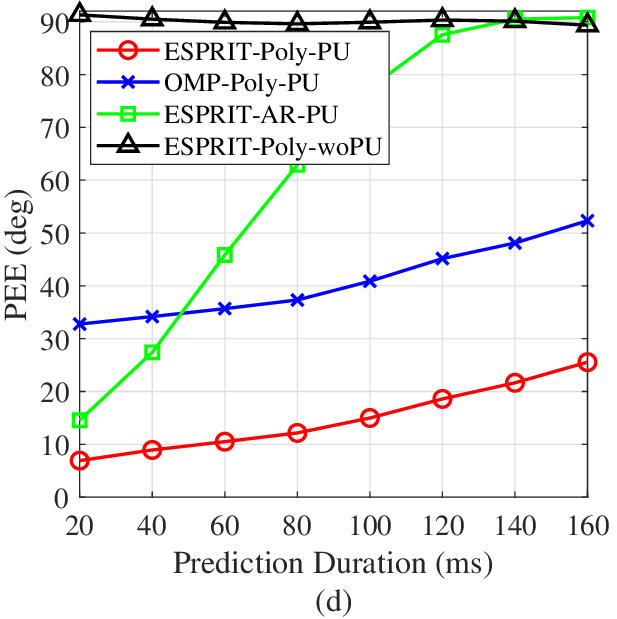}
    \label{fig:PeeVsTime}
  }
  \subfigure{
    \includegraphics[width=0.18\linewidth]{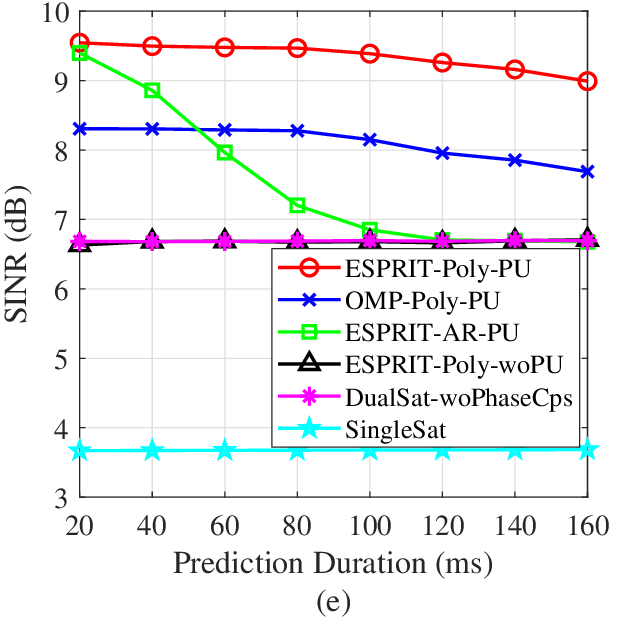}
    \label{fig:SinrVsTime}
  }
  \vspace{-3.5mm}
  \caption{Prediction duration $ T_{\rm pred} $ vs (a) NMSE, (b) TEE, (c) FEE, (d) PEE, (e) SINR. $ P_{\rm beam} = 8 $ dBW, $ M = 12 $, $ T_{\rm period}^{\rm trs} = 20 $ ms.} 
  \label{fig:PerformanceVsTime}
    \vspace{-4mm}
\end{figure*} 

\begin{figure*}[t]
  \centering
  \subfigure{
    \includegraphics[width=0.18\linewidth]{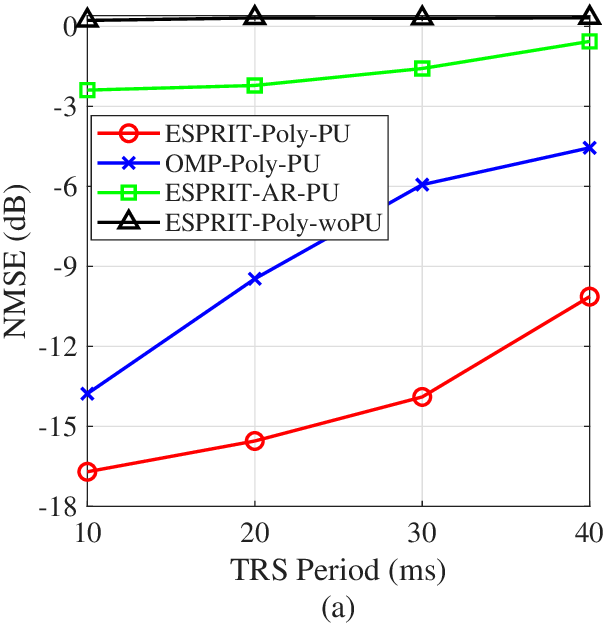}
    \label{fig:NmseVsPeriod}
  }
  \subfigure{
    \includegraphics[width=0.18\linewidth]{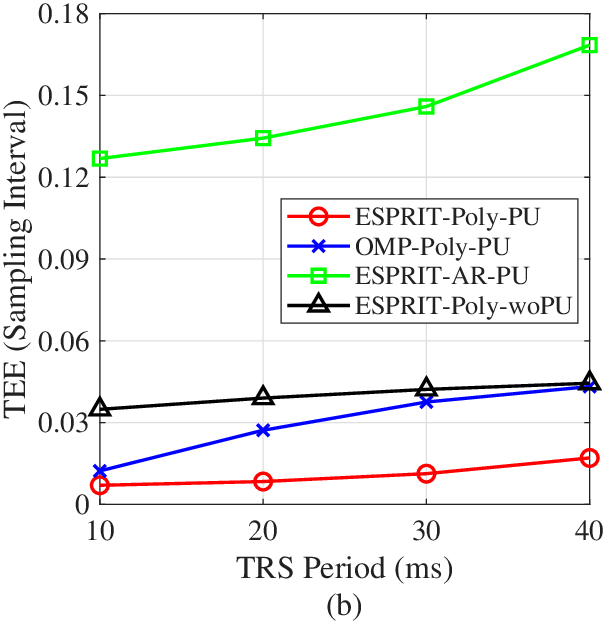}
    \label{fig:TeeVsPeriod}
  }
  \subfigure{
    \includegraphics[width=0.18\linewidth]{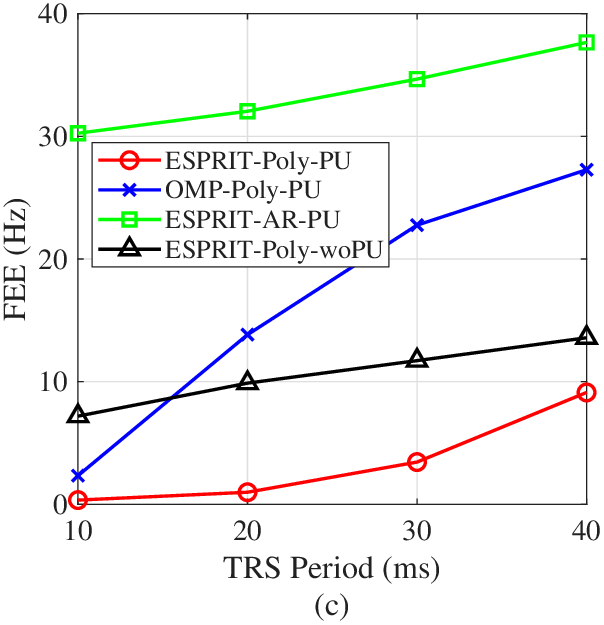}
    \label{fig:FeeVsPeriod}
  }
  \subfigure{
    \includegraphics[width=0.18\linewidth]{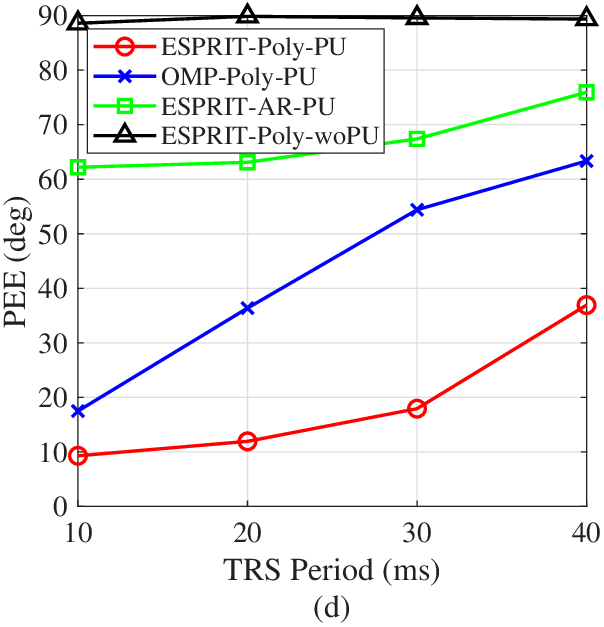}
    \label{fig:PeeVsPeriod}
  }
  \subfigure{
    \includegraphics[width=0.18\linewidth]{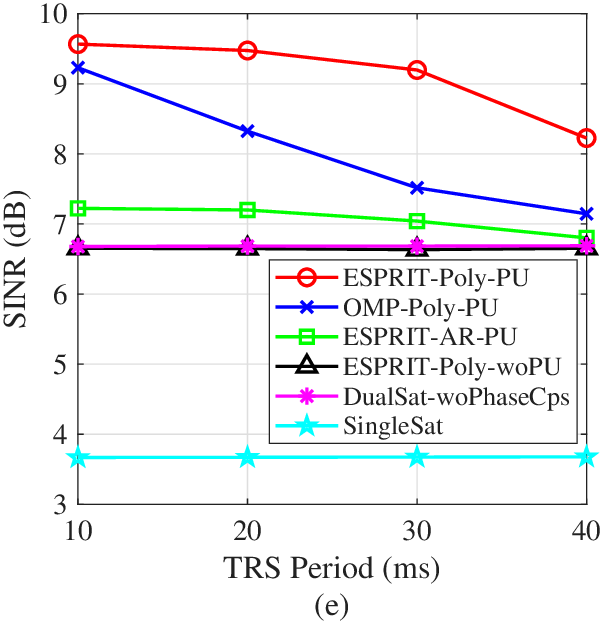}
    \label{fig:SinrVsPeriod}
  }
  \vspace{-3.5mm}
  \caption{TRS period $ T_{\rm period}^{\rm trs} $ vs (a) NMSE, (b) TEE, (c) FEE, (d) PEE, (e) SINR. $ T_{\rm pred} = 80 $ ms, $ P_{\rm beam} = 8 $ dBW, $ M T_{\rm period}^{\rm trs} = 240 $ ms.}
  \label{fig:PerformanceVsPeriod}
  \vspace{-4mm}
\end{figure*} 

\vspace{-2mm}
\subsection{Performance Metric and Baselines} 
The performance of the proposed framework is examined across channel prediction accuracy, TFP-Sync precision, and downlink coherent transmission efficiency. The relevant performance metrics are defined as follows:
\begin{itemize} 

  \item CSI acquisition normalized mean square error~(NMSE): 
  {\small
  \begin{align}
    \mathsf{NMSE}_{p}^{(m)} = \frac{ \sum_{u=0}^{U-1} \sum_{s=0}^{S-1}
    \Vert \mathbf{h}_{sup}^{{\rm trs},(m)} - \hat{\mathbf{h}}_{sup}^{{\rm trs},(m)} \Vert_{2}^{2} }
    { \sum_{u=0}^{U-1} \sum_{s=0}^{S-1} \Vert \mathbf{h}_{sup}^{{\rm trs},(m)} \Vert_{2}^{2} }. 
  \end{align}}

  \item Timing offset estimation error (TEE), frequency offset estimation error (FEE), and phase estimation error (PEE): 
  \vspace{-2mm}
  \begin{subequations}
    \small
    \begin{align}
      &\!\!\!\!\!\mathsf{TEE}_{p}^{(m)} = \frac{\sum_{s=0}^{S-1}\sum_{u=0}^{U-1}|\tau_{sup0}^{(m)} - \tau_{sup}^{{\rm dcps},(m)}|}{SU} , \\
      &\!\!\!\!\!\mathsf{FEE}_{p}^{(m)} = \frac{\sum_{s=0}^{S-1}\sum_{u=0}^{U-1}|\nu_{sup0}^{(m)} - \nu_{sup}^{{\rm dcps},(m)}|}{SU}, \\
      &\!\!\!\!\!\mathsf{PEE}_{p}^{(m)} = \frac{\sum_{s=0}^{S-1}\sum_{u=0}^{U-1} \Vert \angle( \mathbf{h}_{sup}^{{\rm trs},(m)} \odot ( \hat{\mathbf{h}}_{sup}^{{\rm trs},(m)} )^{*} ) \Vert_{1}} {SUN}.
    \end{align}
  \end{subequations}
  
  \item SINR for coherent transmission: 
  {\small
  \begin{align}
    \!\!\!\!\!\mathsf{SINR}_{p}^{(m)} = {\textstyle  \sum_{u=0}^{U-1}} \frac{ \Vert \mathbf{y}_{up}^{{\rm deff},(m)} \Vert_{2}^{2} }{ U ( \Vert \mathbf{z}_{up}^{{\rm int},(m)} \Vert_{2}^{2} + \Vert \mathbf{z}_{up}^{{\rm nse},(m)} \Vert_{2}^{2} ) }.
  \end{align}}
\end{itemize} 

\begin{table}[t]
	\small
	\centering
	\caption{Basic System Parameters \cite{xiang2024massive,38.811,jaeckel20225g,wang2025statistical}}
    \vspace{-1mm}
	\label{tab:system parameters}
	\begin{tabular}{ll}
		\hline
		\textbf{System Parameters} & \textbf{Value}  \\ \hline  
		Carrier frequency $ f_{\text{c}} $ & $2$ GHz \\ 
    FFT size $ N $ & 2048 \\
    CP length $ N_{\text{cp}} $ & $144$ \\
    Subcarrier spacing $ \Delta f $ & $15$ kHz \\
    Number of used subcarriers $N_{\rm sc} $ & $1632$ \\ \hline
    Number of TRS transmissions $ M $ & $12$ \\
    Period of TRS $ T_{\text{period}}^{\text{trs}} $ & $20$ ms \\
    Prediction duration $ T_{\rm pred} $ & $ 80 $ ms \\
    Number of symbols in one TRS $ N_{\text{sym}}^{\text{trs}} $ & $4$ \\
    Symbol indices of one TRS $ \mathcal{N}_{\text{sym}}^{\text{trs}} $ & $\{4, 9, 18, 23\}$ \\
    Number of slots in one TRS $ N_{\rm slot}^{\rm trs} $ & $ 2 $ \\
    Number of transmission combs $ N_{\text{tc}} $ & $4$ \\ \hline
    Number of cooperative satellites $ S $ & $2$ \\
    Per-antenna gain of satellites & $0$ dBi \\ 
    Number of beams per satellite $ B $ & $10$ \\
    Per-beam power of satellites $ P_{\rm beam} $ & $8$ dBW \\
    Number of satellite antennas $ [N_{\rm x},N_{\rm y}] $ & $ [32,32] $ \\
    Gain-to-noise-temperature of UEs & $-33.6$ dB/K \\
    Radius of BPs & $ 8.5 $ km \\ 
    Radius of TBAs & $ 1 $ km \\ 
    Channel model & NTN\_Urban\_LOS \\
    Other loss & $ 4 $ dB\\ \hline
    Constellation Type & Walker-Delta \\
    Orbital altitude $ h_{\rm sat} $ & $350$ km \\
    Orbital Planes & $ 110 $ \\
    Satellites Per Plane & $ 60 $ \\ 
    Inclination Angle & $ 53^{\circ} $ \\ \hline
	\end{tabular}
  \vspace{2mm}
\end{table} 

To demonstrate the superiority of the proposed scheme,
we compare it against the following baselines: 
\begin{itemize}
  \item \textbf{OMP-Poly-PU}: Utilizes the orthogonal matching pursuit (OMP)-based algorithm with a uniform dictionary \cite{determe2015exact} for single-TRS-based channel parameter estimation, followed by cross-TRS PU and polynomial fitting for channel prediction in \secref{sec:channel prediction}. 
  \item \textbf{ESPRIT-AR-PU}: Utilizes the ESPRIT algorithm for single-TRS-based channel parameter estimation, followed by cross-TRS PU in \secref{sec:channel prediction} and autoregressive (AR) modeling \cite{lv2019channel} for channel prediction. 
  \item \textbf{ESPRIT-Poly-woPU}: Utilizes the ESPRIT algorithm for single-TRS-based channel parameter estimation, followed by polynomial fitting for channel prediction in \secref{sec:channel prediction}, without cross-TRS~PU.
\end{itemize} 
The baselines are designed to isolate different components of the proposed framework. Specifically, OMP-Poly-PU evaluates the impact of single-TRS-based channel parameter extraction, ESPRIT-AR-PU assesses the adopted temporal evolution model, and ESPRIT-Poly-woPU highlights the role of cross-TRS PU. Thus, each baseline differs from ESPRIT-Poly-PU in one key component, enabling an interpretable comparison aligned with the algorithmic structure in \secref{sec: JCAFPS}. 
In terms of SINR, the evaluation extends beyond the aforementioned baselines to include single-satellite transmission and dual-satellite cooperative transmission without phase precompensation, denoted as \textbf{SingleSat} and \textbf{DualSat-woPhaseCps}. 

\subsection{Effect of Per-Beam Power} 

\figref{fig:PerformanceVsPt} presents the performance of the proposed and baseline schemes as a function of per-beam transmit power. 
Across the entire power regime, the proposed scheme exhibits superior performance in CSI acquisition, TFP-Sync, and downlink coherent transmission, consistently outperforming the benchmarks. 
The gap to OMP-Poly-PU confirms that limited delay resolution causes leakage-induced estimation distortion, which propagates to synchronization errors. 
The performance loss of ESPRIT-Poly-woPU further shows that accurate single-TRS estimation alone is insufficient without cross-TRS phase unwrapping. 
In addition, the consistent advantage over ESPRIT-AR-PU verifies that the polynomial model better captures the channel evolution in the considered scenario. 
As a result, the proposed scheme enables dual-satellite coherent transmission to approach the theoretical $6$ dB and $3$ dB SINR gains relative to SingleSat and DualSat-woPhaseCps in \figref{fig:SinrVsPt}, respectively.

\subsection{Effect of Prediction Duration} 

Given the trade-off between the prediction duration and feedback overhead, we examine the evolution of key performance metrics over the prediction duration in \figref{fig:PerformanceVsTime}. 
Except for ESPRIT-Poly-woPU, all predictive schemes degrade monotonically as the prediction duration $ T_{\rm pred} $ increases, whereas the proposed method shows the slowest degradation by virtue of the high fidelity of the polynomial model in capturing channel temporal dynamics. 
In particular, at $ T_{\rm pred} = 160 $ ms, the proposed scheme maintains robust performance with NMSE below $ -10 $ dB, TEE below $ 0.01T_{\rm s} $, FEE below $ 2 $ Hz, and PEE below $25^{\circ}$, corresponding to a marginal SINR loss of approximately $6\%$. 
OMP-Poly-PU follows a similar trend but remains limited by estimation leakage, while ESPRIT-AR-PU suffers from recursive error propagation and its SINR collapses to the DualSat-woPhaseCps floor once $ T_{\rm pred} $ reaches $120$ ms, as illustrated in \figref{fig:SinrVsTime}. 
These results confirm that the polynomial-based design is more suitable for long-horizon predictive synchronization. 

\subsection{Effect of TRS Period}
Optimizing the trade-off between performance and pilot overhead constitutes a pivotal aspect of system design. 
In \figref{fig:PerformanceVsPeriod}, we characterize the performance of various schemes as a function of the TRS period. 
It is observed that schemes incorporating PU exhibit a performance decline as $ T_{\rm period}^{\rm trs} $ extends, whereas ESPRIT-Poly-woPU displays a consistently poor performance floor irrespective of $ T_{\rm period}^{\rm trs} $, owing to its inability to recover phase information. 
This performance degradation can be attributed to two factors. First, enlarging the TRS period reduces the temporal density of channel observations, thereby increasing the estimation error of the polynomial coefficients. Second, reliable phase unwrapping requires the TRS period to remain below a critical threshold, which is determined by the achievable frequency offset estimation accuracy, as discussed in \secref{sec:trs design}. 
OMP-based baseline fails beyond $20$ ms due to insufficient frequency offset estimation accuracy for reliable phase unwrapping, which causes severe SINR degradation. By comparison, the proposed scheme is markedly more robust and still achieves at $ T_{\rm period}^{\rm trs}=40$ ms an SINR comparable to that of the best baseline at $20$ ms. This indicates that the proposed design can effectively halve the pilot overhead while preserving superior coherent transmission performance.

\section{Conclusion}
\label{sec:conclusion}

This paper investigated the stringent TFP-Sync challenge in MSCT and developed a closed-loop JCAFPS framework that jointly integrates CSI acquisition, predictive feedback, and UE-specific precompensation. By exploiting deterministic orbital motion and dominant LoS propagation, we established a polynomial channel evolution model and showed that the composite multi-satellite channel exhibits nonlinear time-frequency-varying phase behavior, which fundamentally requires symbol- and subcarrier-wise phase precompensation for coherent transmission. Building on this insight, the proposed framework enables practical predictive synchronization for MSCT with reliable CSI acquisition and feedback. 
Numerical results verified that the proposed framework enables reliable CSI acquisition and stringent TFP-Sync, thereby allowing dual-satellite DP to nearly attain the theoretical $6$ dB gain over single-satellite transmission, with strong robustness to long prediction durations and large TRS periods.





\ifCLASSOPTIONcaptionsoff
  \newpage
\fi



\bibliographystyle{IEEEtran}
\bibliography{IEEEabrv,reference}

\end{document}